\documentclass[aps,pra,onecolumn,amsmath,amssymb,nofootinbib,superscriptaddress]{revtex4-2}
\usepackage{adjustbox}
\usepackage[utf8]{inputenc} 
\usepackage{hyperref}       
\usepackage{url}            
\usepackage{booktabs}       
\usepackage{amsfonts}       
\usepackage{nicefrac}       
\usepackage{microtype}      
\usepackage{natbib}
\usepackage{svg}

\usepackage{bm}
\usepackage{bbm}
\usepackage{braket}
\usepackage{graphicx}
\usepackage[caption=false]{subfig}
\usepackage{float}
\usepackage{amsthm}
\usepackage{algorithmic}
\usepackage[linesnumbered,ruled]{algorithm2e}
\SetKwInOut{Parameter}{parameter}

\usepackage{mathtools}
\usepackage{nccmath}
\usepackage{color}
\usepackage{caption}
\usepackage{titlesec}

\usepackage{soul}

\newcommand{\Tr}{\text{Tr}}

\begin{document}

\title{Multimodal deep representation learning for quantum cross-platform verification}
\author{Yang Qian}
\affiliation{School of Computer Science, Faculty of Engineering, University of Sydney, NSW 2008, Australia
	}
	
\author{Yuxuan Du}
\thanks{duyuxuan123@gmail.com}
\affiliation{JD Explore Academy, Beijing 101111, China}

\author{Zhenliang He}
\affiliation{JD.COM, Beijing 101111, China}
	
\author{Min-Hsiu Hsieh}
\affiliation{
Hon Hai (Foxconn) Research Institute, Taipei 114699, Taiwan
}
		
\author{Dacheng Tao}
\affiliation{School of Computer Science, Faculty of Engineering, University of Sydney, NSW 2008, Australia
	}

\date{\today}


\begin{abstract}
Cross-platform verification, a critical undertaking in the realm of early-stage quantum computing, endeavors to characterize the similarity of two imperfect quantum devices executing identical algorithms, utilizing minimal measurements.  While the random measurement approach has been instrumental in this context, the quasi-exponential computational demand with increasing qubit count hurdles its feasibility in large-qubit scenarios. To bridge this knowledge gap, here we introduce an innovative multimodal learning approach, recognizing that the formalism of data in this task embodies two distinct modalities: measurement outcomes and classical description of compiled circuits on explored quantum devices, both enriched with unique information. Building upon this insight, we devise a multimodal neural network to independently extract knowledge from these modalities, followed by a fusion operation to create a comprehensive data representation. The learned representation can effectively characterize the similarity between the explored quantum devices when executing new quantum algorithms not present in the training data. We evaluate our proposal on platforms featuring diverse noise models, encompassing system sizes up to 50 qubits. The achieved results demonstrate a three-orders-of-magnitude improvement in prediction accuracy compared to the random measurements and offer compelling evidence of the complementary roles played by each modality in cross-platform verification.  These findings pave the way for harnessing the power of multimodal learning to overcome challenges in wider quantum system learning tasks.
\end{abstract}

\maketitle

\section{Introduction} 
 The imperfections inherent in contemporary quantum computers \cite{preskill2018quantum,zhu2022quantum,deng2023gaussian} underscore the need for efficient control, calibration, and validation protocols, a prerequisite to reach their practical utility and advance the development of next-generation quantum computing \cite{kim2023evidence}. In pursuit of validation protocols, cross-platform verification emerges as a pivotal endeavor, primarily concerned with quantifying the similarity between quantum states generated by two quantum computers while executing nominally identical arbitrary circuits \cite{carrasco2021theoretical,suau2021single,anshu2022distributed,knorzer2023cross}. A leading strategy in tackling cross-platform verification is the random measurement approach \cite{elben2020cross,aaronson2018shadow,huang2020predicting,zhu2022cross}. Nevertheless, recent studies have revealed significant challenges of this strategy when applied to large-qubit quantum systems, as the estimation error inversely and quasi-exponentially scales with the required number of measurements \cite{elben2020cross}. In light of these challenges, a fundamental question surfaces: can a solution be found to address cross-platform verification beyond quantum devices comprising tens of qubits?

In this study, we overcome this challenge by harnessing the potential of \textit{multimodal} deep representation learning, underpinned by two fundamental tenets. The first tenet is that the practical interest of cross-platform verification generally revolves around quantifying the similarity between the outputs of two quantum devices as they execute a class of nominally identical circuits drawn from a certain \textit{distribution}. Recognizing this, we transition into a learning paradigm, with the objective of training a learner capable of using a minimal set of measurements to predict the similarity between two devices when operating on new circuits sampled from the same distribution. The second tenet is that cross-platform verification inherently entails two distinct modalities of data: the measurement outcomes and the classical descriptions of the compiled circuits on each quantum device. These modalities carry rich, unique, and complementary information. A learner trained on this multimodal data can discern cross-platform verification with a   deeper insight compared to reliance on a solitary modality.

For concreteness, we devise a novel multimodal deep learning model for quantum cross-platform verification, dubbed \underline{M}easurement-\underline{C}ircuit Driven \underline{N}eural \underline{N}etwork (MC-Net). Conceptually, MC-Net adopts a permutation-invariant neural network and a graph neural network to separately distill knowledge from the modalities of measurement outcomes and circuit descriptions, respectively. Then, a fusion operation is applied to merge the distilled knowledge into a comprehensive and compact data representation. This representation, after training, reveals how factors such as varying system noise, qubit connectivity, and differences in circuit layout influence the similarity of two quantum devices, and the cross-platform verification can be efficiently completed by calculating the cosine similarity of two representations. The knowledge distillation on two modalities enables the superiority of MC-Net in the realm of cross-platform verification compared to prior learning-based methods, centering on the single modality data. As an additional benefit, MC-Net can be directly extended to accomplish purity and entanglement entropy estimation tasks, which may be of independent interest.

To evaluate the efficiency of MC-Net, we focus on a specific class of quantum circuits tailored for cross-platform verification on near-term quantum devices. Envisioned by the fact that most near-term quantum algorithms are inherently hardware-efficient, optimized for effective operation on these devices, the explored class of quantum circuits consists of variational single-qubit and fixed two-qubit gates arranged in diverse configurations within a predefined circuit depth. This circuit category is versatile and encompasses a broad range of near-term quantum algorithms, including quantum neural networks \cite{havlivcek2019supervised,cerezo2021variational,tian2023recent}, variational quantum Eigen-solvers \cite{kandala2017hardware,tang2021qubit}, quantum simulations (Trotter methods or other high-order methods) \cite{smith2019simulating,barratt2021parallel,gibbs2022long,kim2023evidence}, and random circuit sampling \cite{arute2019quantum,zhu2022quantum}. We conduct a systematic analysis on this class of circuits. Numerical results reveal a remarkable improvement in terms of prediction accuracy achieved by our proposal compared to the random measurement approach \cite{zhu2022cross}. Additionally, we conduct an extensive ablation study to demonstrate the benefits of harnessing quantum circuit architecture information in cross-platform verification.

\section{Main Results} 
\begin{figure*}
\centering
\includegraphics[width=0.9\textwidth]{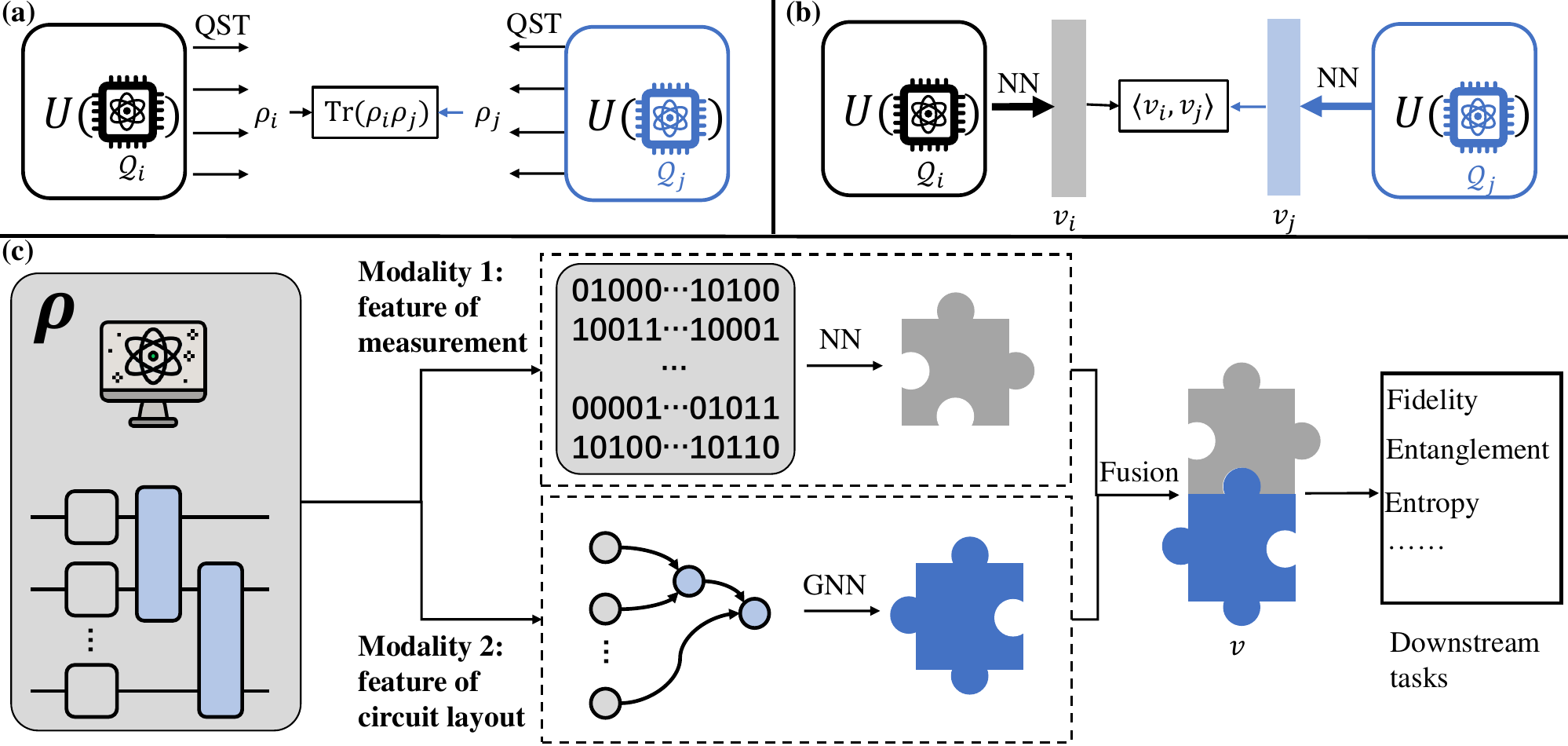}
\caption{\small{\textbf{The framework of MC-Net for cross-platform fidelity estimation.} (a) Conventional methods such as quantum state tomography and random measurements in solving cross-platform fidelity estimation. (b) Instead of (approximately) reconstructing quantum states, MC-Net completes this task by harnessing neural networks to learn the states' low-dimensional and comprehensive representations $\bm{v}_i$ and $\bm{v}_j$ from two modalities, where the cross-platform fidelity is predicted by computing the similarity of the learned representations with $\langle \bm{v}_i, \bm{v}_j\rangle$. (c) The implementation of MC-Net consists of two parts. First, MC-Net employs a permutation-invariant neural network and a graph neural network to separately distill knowledge from the measurement modality and circuit-encoded modality, generating representations for each (denoted as a piece of puzzle). Second, these two feature vectors are fused to create a unified representation of the input state (put the two puzzle pieces together), i.e., $\bm{v}_i$ or $\bm{v}_j$, providing complementarily richer information than individual modalities for downstream tasks.}}
\label{fig:qsv}
\end{figure*}

We first introduce the formal definition of cross-platform verification. Suppose that there are two $N$-qubit devices $\mathcal{Q}_i$ and $\mathcal{Q}_j$ whose noise models are described by $\mathcal{E}_i$ and $\mathcal{E}_j$, respectively. When these two devices receive an identical quantum algorithm described by a unitary $U\in \mathcal{SU}(2^N)$, cross-platform verification concerns the similarity of their output states, named \textit{cross-platform fidelity} \cite{liang2019quantum},  i.e.,    
\begin{equation}\label{eq:cross-fidelity}
    \mathcal{F}(\rho_i,\rho_j) = \frac{\Tr(\rho_i\rho_j)}{\sqrt{\Tr(\rho_i^2)\Tr(\rho_j^2)}} \in [0, 1],
\end{equation}
where $\rho_i= \mathcal{E}_{i}(U\rho_0 U^{\dagger})$, $\rho_j= \mathcal{E}_{j}(U\rho_0 U^{\dagger})$, and $\rho_0=(\ket{0}\bra{0})^{\otimes N}$. Considering that $\mathcal{E}_i$ and $\mathcal{E}_j$ are generally distinct, caused by the variations in hardware architectures, different compiling strategies, qubit configurations, noise levels, and other factors, the quantification of  $\mathcal{F}(\rho_i,\rho_j)$ provides us with deeper insights into hardware-specific errors and the reliability and adaptability of quantum algorithms. As shown in Fig.~\ref{fig:qsv}(a), conventional methods towards this problem include quantum state tomography and random measurements, where the classical description of $\rho_i$ and $\rho_j$ is (approximately) obtained followed by an explicit calculation of $\mathcal{F}(\rho_i,\rho_j)$ \cite{elben2020cross,zhu2022cross}. However, the runtime and memory cost of both methods (quasi) exponentially scale with $N$, which presents a formidable challenge for manipulating the large-qubit scenario.

Besides the infeasibility for the large $N$, there is another deficiency of prior methods in cross-platform verification from a practical perspective. That is, the overarching goal of cross-platform verification revolves estimating the similarity of two devices for a class of quantum circuits rather than a single circuit. In this respect, it is desired to  develop an efficient algorithm minimizing the risk 
\begin{equation}\label{eq:risk}
	\mathbb{E}_{\mathcal{E}_i, \mathcal{E}_j, U\sim \mathbb{U}}[|\mathcal{A}(\bm{x}_i, \bm{x}_j;\bm{w}) -    \mathcal{F}(\rho_i,\rho_j)|],
\end{equation}
where $\mathcal{A}(\bm{x}_i, \bm{x}_j;\bm{w})$ refers to the predicted cross-platform fidelity returned by a classical learner, $\bm{x}_i$ (or $\bm{x}_j$) amounts to the classical description of $\rho_i$ (or $\rho_j$), $\bm{w}$ denotes trainable parameters, and $\mathbb{U}$ and $\mathcal{E}_i$ (or $\mathcal{E}_j$) represents the distribution over quantum algorithms and the noise model on $\mathcal{Q}_i$ (or  $\mathcal{Q}_j$) respectively that will be elucidated later. For instance, when a user deploys a quantum neural network on two platforms for image classification, the flexibility to alter the network layout and update trainable parameters necessitates the evaluation of $ \mathcal{F}(\rho_i,\rho_j)$ with the varied $U\sim \mathbb{U}$.

To compensate for the deficiency of conventional methods, here we propose a multimodal learning method, dubbed Measurement-Circuit Driven Neural Network (MC-Net), to address the cross-platform verification problem in Eq.~(\ref{eq:risk}) when $N$ becomes large. The proposed MC-Net falls into the supervised learning framework and comprises three fundamental components: multimodal dataset collection,  model implementation, and model optimization. In the following, we briefly recap the key insight and realization of these three components, and defer the omitted details to  SM~\ref{app:prepare-state} and \ref{app:mc-net}.  

The multimodal dataset used for training MC-Net, denoted by $\mathcal{D}_{\text{Tr}}=\{(\bm{x}_i^{(s)},\bm{x}_j^{(s)},\mathcal{F}^{(s)}_{ij})\}_{s=1}^S$ consists of $S$ samples randomly drawn from  $\mathbb{U}$, where $\bm{x}^{(s)}_i$ (or $\bm{x}^{(s)}_j$) represents data features of the $s$-th sample for $\mathcal{Q}_i$ (or $\mathcal{Q}_j$), and the label $\mathcal{F}^{(s)}_{ij}=\mathcal{F}(\rho_i,\rho_j)$ corresponds to the cross-platform fidelity defined in Eq.~(\ref{eq:cross-fidelity}). A central challenge in  constructing $\mathcal{D}_{\text{Tr}}$ is identifying the specific information to retain from an exponentially large state space, i.e., the data features must strike a balance, encompassing adequate information for the task while maintaining memory and computational efficiency. To address this challenge, we design a multimodal strategy to engineer the data features, where each sample contains two parts: the measurement modality and circuit modality. Particularly, the measurement modality is gathered by conducting the classical shadow of the quantum state $\rho^{(s)}$ from $M$ random Pauli measurements \cite{huang2020predicting}, which warrants a succinct classical representation of a quantum state. The circuit modality is obtained by encoding the circuit layout and the device configuration from $U^{(s)}$ and $\mathcal{E}^{(s)}$ as a directed acyclic graph (DAG). The handcrafted multimodal data features contain a more comprehensive and accurate description of a quantum system compared to the single modality features,  enabling MC-Net to use fewer training samples and measurements to reach a satisfied prediction accuracy.  See SM~\ref{app:mc-net} for detailed construction of $\mathcal{D}_{\text{Tr}}$.

Before moving on to present the implementation details of MC-Net, let us emphasize the intrinsic hardness in the design of a multimodal deep learning model for handling the dataset $\mathcal{D}_{\text{Tr}}$. Note that the original formats for both the measurement and circuit modalities in $\mathcal{D}_{\text{Tr}}$ are incompatible with neural networks. Accordingly, a major challenge is to determine how to effectively preprocess these raw data features, making them conducive to integration within the chosen neural network architecture. Furthermore, the process of leveraging prior knowledge to enhance the efficiency of the learning model for extracting knowledge from multimodal features remains a largely uncharted realm. Additionally, the means to synergize the extracted knowledge from diverse modalities to bolster cross-platform verification has yet to be explored.

To conquer the aforementioned challenges, we devise the architecture illustrated in Fig.~\ref{fig:qsv}(c) to implement MC-Net. In particular, the measurement modality is processed by a permutation-invariant neural network similar to PointNet \cite{qi2017pointnet} to output a feature vector related to the measurements. The permutation-invariance allows MC-Net to well respect the inherent disorder of measurement data obtained from multiple random snapshots. Besides, the circuit modality is processed by a graph neural network to obtain a feature vector related to the circuit-induced DAG. These two feature vectors are aggregated by a fusion operation using bilinear pooling \cite{kim2016hadamard}, ultimately yielding a unified, comprehensive, and low-dimensional representation $\bm{v}$. Denote the distilled representation of MC-Net for the $i$-th ($j$-th) quantum device on the $s$-th sample as $\bm{v}_i^{(s)}$ ($\bm{v}_j^{(s)}$), the estimated cross-platform fidelity amounts to the cosine similarity of these two representations, i.e.,
\begin{equation}\label{eq:ml}
    \hat{\mathcal{F}}_{ij}^{(s)} = \frac{\left\langle\bm{v}_i^{(s)}, \bm{v}_j^{(s)}\right\rangle}{\left\|\bm{v}_i^{(s)}\right\|_2\left\|\bm{v}_j^{(s)}\right\|_2}.
\end{equation}

Given the assembled multimodal training dataset $\mathcal{D}_{\text{Tr}}$, MC-Net utilizes it to optimize the trainable parameters $\bm{w}$ to minimize the objective function 
\begin{equation}\label{eq:loss-func}
\epsilon(\bm{w})=\frac{1}{S}\sum_{s=1}^S \mathrm{D}\left(\hat{\mathcal{F}}_{ij}^{(s)}, \mathcal{F}^{(s)}_{ij}\right),	
\end{equation}
where $\mathrm{D}(\cdot, \cdot)$ is the per-sample loss, e.g., the mean squared error, and $\hat{\mathcal{F}}_{ij} = \mathcal{A}(\bm{x}_i^{(s)},\bm{x}_j^{(s)},\bm{w})$ refers to the predicted cross-platform fidelity returned by MC-Net. The optimization of $\bm{w}$ is completed by the gradient-descent methods. 

During inference, once we have the profile $(\bm{x}_i, \bm{x}_j)$ of an unknown pair of states $(\rho_i,\rho_j)$ prepared by the same $U$ sampled from $\mathbb{U}$, we can efficiently estimate the cross-platform fidelity by performing a fast feedforward through the optimized MC-Net. MC-Net's ability of efficiently generalizing to new states enables the assessment of state consistency on different platforms after changing the preparation circuit, e.g., updating the parameters of variational quantum circuits, in a computationally friendly manner.

\noindent\textbf{Remark}. MC-Net breaks down the procedure of cross-platform fidelity estimation into two subroutines: the multimodal representation learning of a quantum state and the similarity calculation. This decoupling allows for easy adaptation of MC-Net to various tasks. Namely, by replacing the task-related head responsible for cross-platform verification with other tasks, such as state purity and entanglement, MC-Net can be repurposed to address different aspects of quantum system analysis \cite{man2011homodyne,horodecki2003measuring,acharya2019measuring}. Additionally, this decoupling  design facilitates extending MC-Net to verify multiple devices $\{\mathcal{Q}_k\}$ simultaneously, a feature that will be demonstrated in the subsequent experiments. See SM~\ref{app:extension} for elaborations.

\begin{figure*} 
\centering
\includegraphics[width=1.0\textwidth]{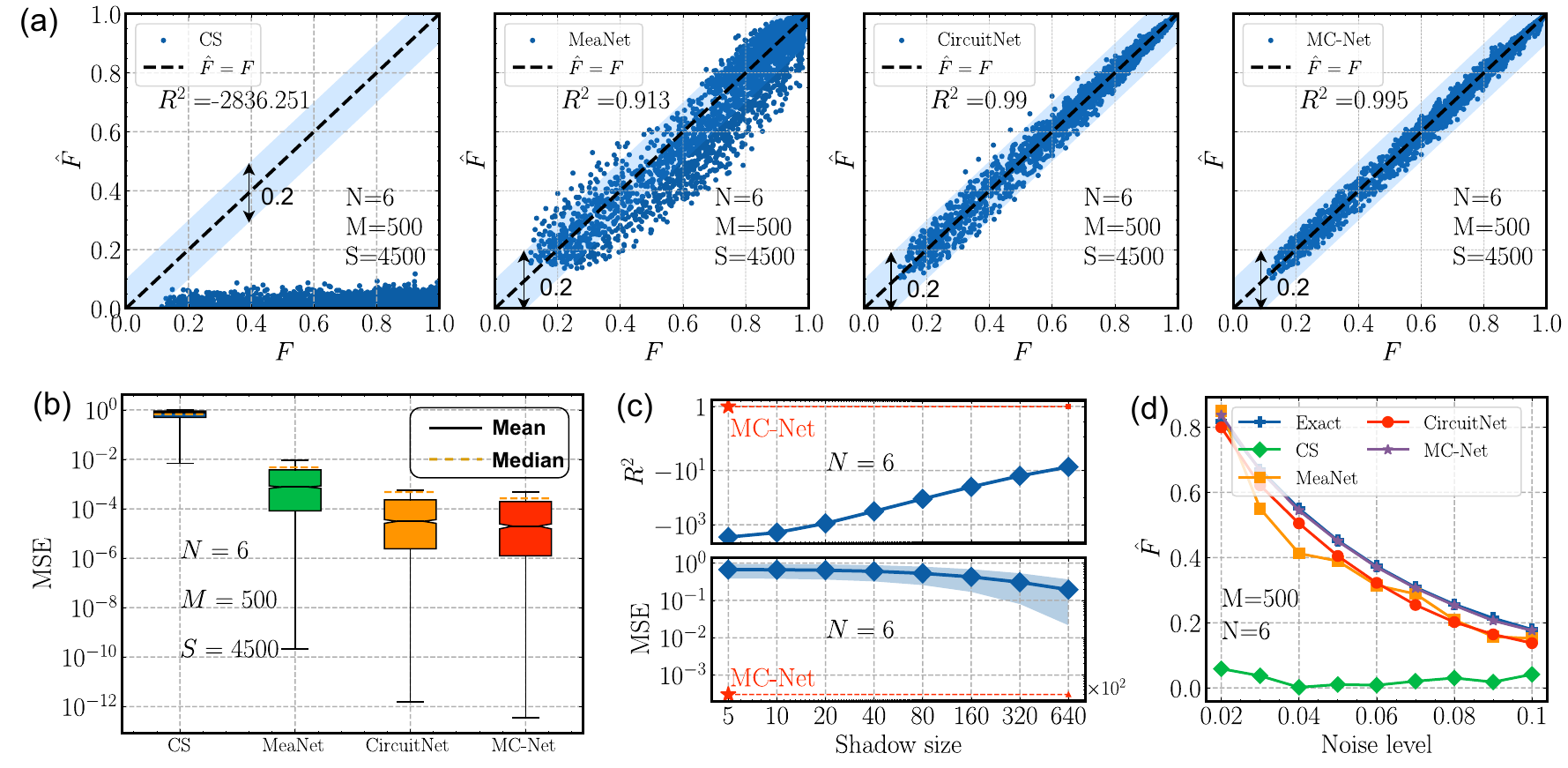}
\caption{\small{\textbf{Cross-platform fidelity estimated by classical shadows (abbreviated as `CS'), the single measurement branch MeaNet, the single architecture branch CircuitNet and the multi-modal model MC-Net for $6$-qubit quantum systems.} (a) The correlation between ground truth cross-platform fidelity and fidelity predicted by four approaches with $500$ random Pauli measurements and $100$ circuits. $R^2$ represents the coefficient of determination. The black solid line $y=x$ refers to perfect prediction. (b) The mean square error (MSE) comparison by four approaches. The labels `Mean' and `Std' represent the arithmetic mean and standard deviation of MSE respectively. (c) The scaling behavior of classical shadow based cross-platform fidelity estimation as a function of the shadow size (number of measurements). The top and bottom panels denote the $R^2$ and MSE respectively. The red star represents the performance of MC-Net. (d) The behaviors of four approaches concerning various noise levels. 
}}
\label{fig:hea_q6}
\end{figure*}
\section{Numerical Simulations} 
To assess our proposal against prior methods, we compare classical shadow-based cross-platform verification (CS) \cite{zhu2022cross} with MC-Net under diverse settings. Moreover, to highlight the benefits of multimodal representation learning, in some experiments, we conduct the ablation study, completed by benchmarking the performance of MC-Net when only a single branch is preserved, dubbed MC-Net's measurement branch (MeaNet) and MC-Net's circuit branch (CircuitNet).

We initially conduct cross-platform verification on platforms simulated with two-qubit depolarizing noise, which have the worst effect on the state \cite{preskill2015lecture}. The explicit form of the explored class of circuits takes a hardware-efficient characteristic, i.e., 
\begin{equation}\label{eqn:circuit-task-1}
	 \left\{U =\prod_{l=1}^L\prod_{n=1}^N e^{-i\theta_{l,n}H_{l,n}}V_l \Big| \theta_{l,n} \sim \text{Unif}(0,2\pi) \right\},
\end{equation}
where   $H_l$ is a random Pauli operator and $V_l\sim \mathbb{V}$ is an entanglement block consisting of multiple CNOT gates connecting physically neighboring qubits. Both the number of CNOT gates and the index of their control qubit are generated randomly (See SM~\ref{app:cir} for details). When creating the training dataset, we initially sample $100$ circuits from Eq.~(\ref{eqn:circuit-task-1}) randomly. For each circuit, we introduce $10$ levels of depolarizing noise, uniformly sampled in the range $[0.01, 0.1]$.  Subsequently, we create pairs of circuits with two different noise strengths applied to the same circuit, resulting in a dataset with a size of $S=4500$. The process of constructing the test dataset follows a similar procedure. The circuits are sampled from the same distribution as those in the training dataset, but they have different layouts compared to the training data.

We evaluate the model's performance using two standard metrics: the coefficient of determination $R^2$ and Mean Square Error (MSE), i.e., $R^2=1-\sum_s(\mathcal{F}^{(s)}-\hat{\mathcal{F}}^{(s)})^2/\sum_s(\mathcal{F}^{(s)}-\Bar{\mathcal{F}})^2$, where $\Bar{\mathcal{F}}$ is the average fidelity over all samples; $\text{MSE}=\frac{1}{S}\sum_s(\mathcal{F}^{(s)}-\hat{\mathcal{F}}^{(s)})^2$. Intuitively, $R^2$ measures how well the model's estimations match the actual values, with a perfect fit having an $R^2$ value of 1, and MSE quantifies the average squared difference between our model's predictions and the actual values.

The overall performance of CS, MC-Net, CircuitNet, and MeaNet on $6$-qubit systems with $M=500$   is illustrated in Fig.~\ref{fig:hea_q6}(a-b). With such a limited number of measurements, CS exhibits diminished accuracy in estimating cross-platform fidelity, as evidenced by its low $R^2$ score of $-2836$ and high MSE of $0.6727$. In contrast, MeaNet, CircuitNet, and MC-Net consistently demonstrate the satisfied fidelity prediction accuracy, boosting $R^2$ values exceeding $0.9$ and MSE values below $0.005$. Among them, MC-Net outperforms the rest, achieving the highest $R^2$ score of $0.995$ and the smallest MSE of $0.0003$, a three-orders-of-magnitude enhancement in terms of MSE precision than CS. 

We further exploit the resource efficiency of MC-Net in comparison to CS in terms of the snapshot $M$. As illustrated in Fig.~\ref{fig:hea_q6}(c), when exponentially increasing shot number $M$ from $500$ to $64000$, the resulting $R^2$ of CS increases from $-2836$ to $-7$, and the resulting MSE of CS experiences approximate linear reduction from $0.67$ to $0.2$. Such accuracy still remains impractical for real-world applications. Nevertheless, MC-Net with only $500$ measurements achieves an extremely low MSE of $10^{-4}$ and a high $R^2$ of $0.995$, which would require over $100,000$ measurements for CS to match.

We subsequently dive into the contribution of measurement modality and circuit modality to the final prediction by zooming in on a single circuit prepared with various noise strengths. We take the state prepared by the circuit with a noise level of $0.01$ as the reference state and calculate the fidelity between the reference state and states prepared by the same circuit with the noise level growing from $0.02$ to $0.1$. The behaviors of four approaches are shown in Fig.~\ref{fig:hea_q6}(d). While the fidelity predicted by MeaNet and CircuitNet exhibits an apparent gap compared to the exact fidelity, the behavior of their combination, MC-Net, is almost identical to the exact value. Although previous results suggest that CircuitNet statistically outperforms MeaNet, Fig.~\ref{fig:hea_q6}(d) reveals that there are cases where MeaNet performs better than CircuitNet, such as for devices with a two-qubit depolarizing noise level of $0.07$ and $0.1$. This phenomenon indicates the complementary nature of measurement modality and circuit modality in enhancing fidelity estimation. Taken together, they contribute to a more accurate prediction of fidelity in quantum systems.

\begin{figure*}
\centering
\includegraphics[width=1.0\textwidth]{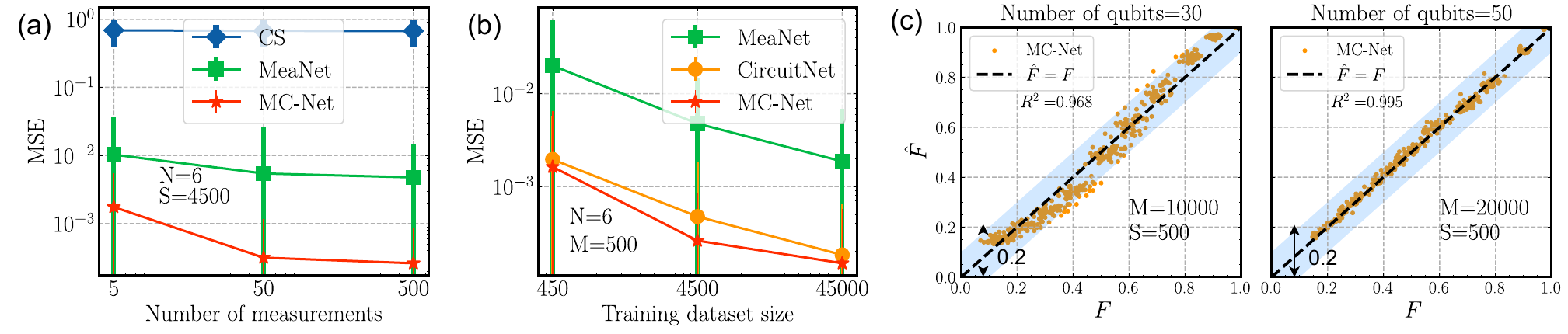}
\caption{\small{\textbf{The performance of the proposed model concerning the training dataset size, the number of measurements, and the number of qubits.} (a) The sensitivity of MSE achieved by CS, MeaNet, and MC-Net to the reduction of measurements. The CircuitNet is ignored in this context because it requires no measurement. (b) The sensitivity of MSE achieved by MeaNet, CircuitNet, and MC-Net to the reduction of training data. (c) The performance of MC-Net when handling larger-scale systems, including a $30$-qubit platform and a $50$-qubit platform.}}
\label{fig:hea_scale}
\end{figure*}

We next examine how MC-Net behaves with the varied number of measurements and the size of the training dataset. As depicted in Fig.~\ref{fig:hea_scale}(a), fixing the number of circuits in the training data $S=4500$, increasing the number of measurements $M$ from $5$ to $500$ leads to a substantial reduction of MSE from $10^{-3}$ to $10^{-4}$ for MC-Net. In contrast, MeaNet shows a slower decrease in MSE, from $10^{-2}$ to $5\times10^{-3}$, and CS exhibits a only marginal change, with the value remaining at the level of $10^{-1}$. Fig.~\ref{fig:hea_scale}(b) underscores the minimal training data requirements of our proposed methods with fixed $M=500$ measurements. By exponentially decreasing the training dataset size $S$ from $4500$ to $450$, MC-Net witnesses a one-order-of-magnitude increase in MSE but still attains the same low MSE level as MeaNet trained on $10^3$ circuits. In contrast, when trained on just $10$ samples, MeaNet struggles to predict cross-platform fidelity, resulting in a significantly higher MSE of $10^{-2}$. MC-Net's ability to deliver valid estimations with a small training dataset size and a modest number of measurements makes it a viable choice for assessing large-scale quantum platforms.

To show the compatibility of our proposed model as the system size increases, we then expand our experiments to encompass larger-scale quantum systems, specifically involving $30$ qubits and $50$ qubits. It is important to note that the dimensions of both modalities constructed in our model scale polynomially with the system size, alleviating any undue burden on the requisite number of trainable parameters within MC-Net. The noise simulation for these experiments is the same as in the experiments described earlier, with the depolarizing channel strength sampled from a range of smaller values $[10^{-4},10^{-2}]$. Considering the high complexity of large-scale systems, we set the number of measurements $M$ to $10000$ for the $30$-qubit system and $20000$ for the $50$-qubit system. Fig.~\ref{fig:hea_scale}(c) shows the predictions of MC-Net on circuits randomly sampled from Eq.~(\ref{eqn:circuit-task-1}) after training on nearly $100$ circuits following the same distribution. Although 20,000 shots fall short of achieving precise fidelity estimation for large-scale systems using CS, it is evident that the estimated cross-platform fidelity returned by MC-Net still exhibits a strong correlation with the ground truth, with $R^2$ greater than $0.95$. See SM~\ref{app:scale-qubit} for more results about the scaling behavior of MC-Net with system size.
 
\begin{figure}[htp]
\centering
\includegraphics[width=0.48\textwidth]{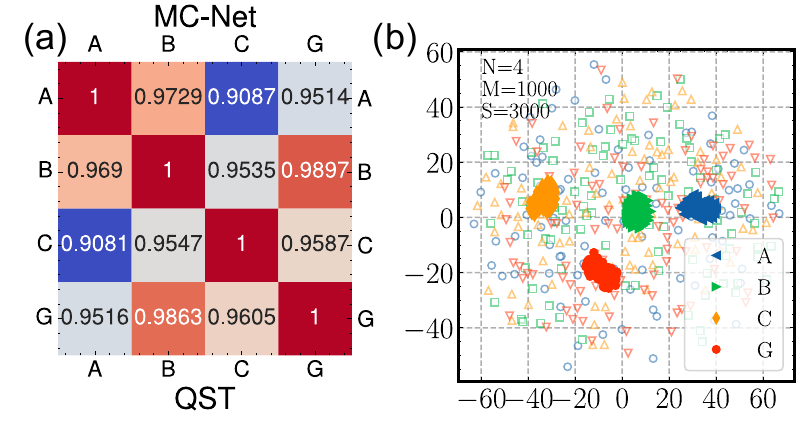}
\caption{\small{\textbf{The performance of MC-Net for platforms with real device noise models.} (a) The cross-platform fidelity between four devices estimated by quantum state tomography and proposed MC-Net. The devices `A', `B', `C' and `G' represent \textit{fake\_almaden}, \textit{fake\_boeblingen}, \textit{fake\_cambridge}, and \textit{fake\_guadalupe} respectively. Fidelity values in the heatmap are obtained by averaging on $100$ random circuits. The lower 
triangle records the results of QST which refers to quantum state tomography, and the upper triangle records the results of MC-Net. (b) The t-SNE visualization of the low-dimensional representation predicted by MC-Net.}}
\label{fig:hea_noisy}
\end{figure}                                               

We last assess the performance of MC-Net on real near-term devices. To do so, we conducted cross-platform verification using MC-Net on four offline platforms that mimic the behaviors of the IBM Quantum system \cite{IBMquantum}. These platforms, namely \textit{fake\_almaden} (A), \textit{fake\_boeblingen} (B), \textit{fake\_cambridge} (C), and \textit{fake\_guadalupe} (G) \cite{Qiskit}, exhibit variations in qubit connection maps, basis gates, and noise configurations. To ensure comprehensive consideration of the quantum platform's characteristics, each hardware-efficient circuit is compiled to match the target platform's specified basis gates and qubit connectivity. Subsequently, we transform the compiled circuit into a DAG. In addition, we incorporated calibration data for each physical qubit, such as T1, T2, and state-preparation-and-measurement error, into our model, encoding them as feature vectors for each graph node. We sample $S=500$ and $S=100$ random $4$-qubit circuits from the same distribution as in previous experiments and execute them on the four platforms mentioned above to construct the training and test datasets.

The simulation results are illustrated in Fig.~\ref{fig:hea_noisy}(a-b). The lower triangle in subplot (a)  represents outcomes obtained via quantum state tomography with 10,000 measurements. In comparison, the upper triangle displays results generated by our model using 1000 random Pauli measurements for each circuit. Remarkably, MC-Net adeptly discerns disparities across different physical devices, achieving nearly equivalent cross-platform fidelity estimations as quantum state tomography. The average estimation error between MC-Net and QST remains consistently at the level of $10^{-4}$. Furthermore, Fig.~\ref{fig:hea_noisy}(b) employs the t-distributed stochastic neighbor embedding (t-SNE) technique \cite{van2008visualizing} to visualize original the measurement features and the low-dimensional representations of all test states predicted by MC-Net. The original features based on random measurements, denoted by the hollow markers, are uniformly distributed, showing no device-specific clues. In contrast, the representations learned by MC-Net, denoted by solid markers, reveal distinct clusters for states prepared on different devices, with no overlap among these clusters. This observation underscores MC-Net's ability to effectively extract features from mixed states, enabling the discrimination of various quantum platforms.

\section{Discussion and Outlook}

In this work, we present a multimodal representation learning framework for quantum cross-platform verification, emphasizing the complementary roles of measurement and circuit modalities to enhance prediction performance. The proposed MC-Net contributes to the fast evaluation of the reliability and adaptability of quantum algorithms across different near-term quantum platforms. Numerical results up to 50 qubits validate the efficacy of MC-Net.

Our work stands as a catalyst for further study on multimodal learning for quantum tasks. There is an opportunity to explore new modalities that provide more comprehensive information about quantum systems, thus enhancing the model's representation accuracy. In addition to the measurement modality and circuit modality used in our model, the geometrical structure of the Hamiltonians can be considered as another modality, which proves essential in predicting the ground state properties of a quantum many-body system \cite{lewis2023improved}. Furthermore, the concrete format of each modality can be updated to improve the model's performance for specific tasks. For instance, the random Pauli measurements employed in our work could potentially be substituted with more advanced strategies \cite{huang2022learning,nguyen2022optimizing,hu2023classical,bu2022classical}. Besides upgrading the dataset, we can introduce more powerful neural networks and advanced fusion strategies that have been proven successful in classical tasks, such as the Transformer \cite{vaswani2017attention}, to handle larger quantum datasets more efficiently. These refinements could open up new avenues for research in quantum system learning.


\newpage 	
\renewcommand{\thefigure}{M\arabic{figure}}	
\setcounter{figure}{0}

\newpage   
\clearpage 
\appendix 
\onecolumngrid
\tableofcontents
 
The outline of supplementary materials is structured as follows. In Sec.~\ref{app:related-work}, we provide a review of prior literature related to our work. Sec.~\ref{app:prepare-state} introduces the construction of a class of quantum circuits and quantum noise models employed in our experiments. Sec.~\ref{app:mc-net} elaborates on the details of MC-Net. In Sec.~\ref{app:extension}, we explore several extensions of MC-Net for new tasks. Finally, Sec.~\ref{app:more-results} presents additional numerical results obtained with MC-Net.

\section{Related work}\label{app:related-work}
In this section, we will review the prior literature related to our work and elucidate the connections and distinctions between our proposed model and existing methods. Initially, we will clarify two random measurement based protocols for cross-platform verification. Following this, a concise survey of the field of artificial intelligence driven quantum system learning is provided.  

\subsection{Random measurement based protocols for cross-platform verification}\label{app:rm}

Random measurements offer a practical solution in approximately expressing a multi-qubit quantum state by an efficient classical representation, which can be subsequently processed on classical computers to predict many properties of the state  \cite{van2012measuring,elben2018renyi,huang2020predicting,knips2020multipartite,ketterer2019characterizing,zhu2022cross,elben2020cross}. Refer to Refs.~\cite{elben2023randomized,huang2022learning} for a comprehensive survey of fundamental theory of random measurements and the relevant applications. In this context, two studies particularly pertinent to our work are protocols grounded in cross correlations and classical shadow approaches, specifically designed for the assessment of cross-platform fidelity.

\medskip
\noindent\textbf{Cross correlations based protocol \cite{elben2020cross}.} The fundamental principle of this protocol is leveraging statistical correlations between random measurement results on distinct devices to infer the overlap of two states. Following notations in the main text, two mixed $N$-qubit states $\rho_i$ and $\rho_j$ are prepared on two quantum devices $\mathcal{Q}_i$ and $\mathcal{Q}_j$. The cross correlations based protocol first applies the same random unitary $U$ to both $\rho_i$ and $\rho_j$ and then performs projective measurements in the computational basis $\ket{\bm{b}}=\ket{b_1,...,b_N}$, where $\bm{b}$ represents a bit string of possible binary measurement outcomes. Then, after multiple measurements, the estimation of the probabilities $P_U^{i}(\bm{b})=\Tr (U\rho_{i}U^\dagger\ket{\bm{b}}\bra{\bm{b}})$ and $P_U^{j}(\bm{b})=\Tr (U\rho_{j}U^\dagger\ket{\bm{b}}\bra{\bm{b}})$ can be acquired by using the collected frequencies. Through randomly sampling different unitaries $U$ and repeating the above procedure, the overlap $\Tr(\rho_i\rho_j)$ in Eq.~(\ref{eq:cross-fidelity}) is calculated by the second-order cross-correlations between $\mathcal{Q}_i$ and $\mathcal{Q}_j$, formulated as
\begin{equation}\label{eq:rm-cc}
    \Tr(\rho_i\rho_j)=2^N\sum_{\bm{b},\bm{b}'}(-2)^{-D[\bm{b},\bm{b}']}\overline{P_U^{i}(\bm{b})P_U^{j}(\bm{b}')},
\end{equation}
where the symbol $\overline{\cdot\cdot\cdot}$ denotes the average over random unitaries $U$, and $D(\bm{b},\bm{b}')$ defines the Hamming distance between bit strings $\bm{b}$ and $\bm{b}'$. The denominator of Eq.~(\ref{eq:cross-fidelity}), i.e., $\Tr(\rho_i^2)$ and $\Tr(\rho_j^2)$, can be inferred by assigning the same subscript in Eq.~(\ref{eq:rm-cc}).

The estimation error of cross-platform fidelity in this protocol arises from two parts: the finite number of random unitaries $N_U$ used to infer the overlap in Eq.~(\ref{eq:rm-cc}) and the finite number of measurements performed for each unitary $N_M$ used to infer the probability $P_U$. An empirical study conducted by Ref.~\cite{elben2020cross} discovered that the required number of measurement budget $N_UN_M$ needed to reduce the statistical error below a fixed value exhibits exponential scaling with the system size $N$, i.e., 
\begin{equation}
	N_UN_M\sim 2^{bN},~\text{with}~b\lesssim 1.
\end{equation}
This scaling behavior highlights the expensive or even unaffordable computational demands for larger quantum systems when aiming to achieve a certain level of fidelity estimation accuracy.

\medskip
\noindent\textbf{Classical shadows based protocol \cite{huang2020predicting}.} This protocol targets reconstructing a quantum state from random measurement results in a memory and computation efficient manner. For clarity, we first present the mechanism of classical shadow and then elucidate the implementation of classical shadows based protocol.  

The classical shadow is a classical description of a quantum state derived from finite random measurements, serving for efficiently estimating the expectation values of certain observables of the state. The procedure of obtaining the classical shadow is as follows. For an $N$-qubit state $\rho$ prepared on a quantum device $\mathcal{Q}$, a unitary $U$ randomly sampled from a fixed ensemble $\mathbb{U}$ is applied on this state and then measured in the computational basis to obtain the binary measurement outcomes $\ket{\bm{b}}=\ket{b_1,...,b_N}$. Based on the measurement results, a classical description of the reverse operation $U^\dagger\ket{\bm{b}}\bra{\bm{b}}U$ is stored in classical memory, which is called the snapshot of the state. The expectation of the snapshot over the randomness of the unitary $U$ and measurement outcome $\bm{b}$ can be viewed as a quantum channel $\mathcal{M}$ mapping $\rho$ to its snapshot:
\begin{equation}
    \mathbb{E}[U^\dagger\ket{\bm{b}}\bra{\bm{b}}U]=\mathcal{M}(\rho).
\end{equation}
If the ensemble of unitaries is a tomographically complete set of measurements, we can recover the state $\rho$ by inverting the channel
\begin{equation}
    \rho = \mathbb{E}[\mathcal{M}^{-1}(U^\dagger\ket{\bm{b}}\bra{\bm{b}}U)].
\end{equation}
Performing the random measurements for $M$ times, the state $\rho$ can be approximated by taking the statistical average:
\begin{equation}
    \hat{\rho}=\frac{1}{M}\sum_{m=1}^M\mathcal{M}^{-1}(U_m^\dagger\ket{\bm{b}_m}\bra{\bm{b}_m}U_m).
\end{equation}

A general choice of the random unitary $U$ is the single-qubit Pauli group $\{U=\otimes_{n=1}^NU_n\big|U_n\in\{\sigma_I,\sigma_X,\sigma_Y,\sigma_Z\}\}$, which is conveniently implemented and highly efficient as suggested in \cite{huang2020predicting}. With this configuration, the classical shadow of the density matrix $\rho$ is expressed as
\begin{equation}
    \hat{\rho}=\frac{1}{M}\sum_{m=1}^M\otimes_{n=1}^N(3U^\dagger_{m,n}\ket{b_{m,n}}\bra{b_{m,n}}U_{m,n}-\mathbbm{I}),
\end{equation}
where $U_{m,n}$ and $b_{m,n}$ represents the selected unitary and measurement outcome for the $n$-th qubit in the $m$-the measurement.

We next introduce the classical shadow based protocol for cross-platform verification proposed by \cite{zhu2022cross}.
Once the classical shadow $\hat{\rho}_i$ and $\hat{\rho}_j$ are obtained for states $\rho_i$ and $\rho_j$ respectively, the cross-platform fidelity estimation can be calculated by
\begin{equation}\label{eq:cs}
    \hat{\mathcal{F}}(\rho_i,\rho_j) = \frac{\Tr(\hat{\rho}_i\hat{\rho}_j)}{\sqrt{\Tr(\hat{\rho}_i\hat{\rho}_i)\Tr(\hat{\rho}_j\hat{\rho}_j)}}.
\end{equation}
With increasing the number of measurements $M$, the error between $\hat{\mathcal{F}}$ and $\mathcal{F}$ in Eq.~(\ref{eq:cross-fidelity}) gradually diminishes and tends towards zero. When the reconstructed state $\hat{\rho}$ is too large to store classically for a large-scale quantum system, the Pauli-based random measurements allow us to employ a memory-efficient version of Eq.~(\ref{eq:cs}) to estimate the fidelity, i.e.,
\begin{equation}
    \hat{\mathcal{F}}(\rho_i,\rho_j)=\frac{\sum_{m=1,m'=1}^{M}\prod_{n=1}^N\Tr(\hat{\rho}_{i,m,n}\hat{\rho}_{j,m',n})}{\sqrt{\sum_{m=1,m'=1}^{M}\prod_{n=1}^N\Tr(\hat{\rho}_{i,m,n}\hat{\rho}_{i,m',n})*\sum_{m=1,m'=1}^{M}\prod_{n=1}^N\Tr(\hat{\rho}_{j,m,n}\hat{\rho}_{j,m',n})}},
\end{equation}
where $\hat{\rho}_{i,m,n}=3U_{m,n}^\dagger\ket{b_{m,n}}\bra{b_{m,n}}U_{m,n}-\mathbbm{I}$ denotes a local classical shadow of the whole state on the $n$-th qubit in system $\mathcal{Q}_i$ when conducting the $m$-th measurement. The computational complexity of this classical shadow-based cross-platform fidelity estimation also scales exponentially in the system's size \cite{huang2020predicting,zhu2022cross}.

\medskip
\noindent\textbf{Remark}. Our work complements and extends the existing literature in this domain, offering broader applications and a more efficient approach. First, one of the modalities employed in MC-Net is the classical shadow of a quantum state. This implies that an advanced version of classical shadow can also enhance our model. Simultaneously, our model can effectively address the versatile properties supported by classical shadow techniques. Second, MC-Net incorporates additional modalities, such as circuit information, to complement the classical shadow. This results in a more efficient understanding of quantum systems and a reduction of the required number of measurements.

\subsection{AI-driven quantum system learning}

Artificial intelligence (AI), known for its powerful ability of distilling useful knowledge from complicated data, is playing an increasingly significant role in expediting traditionally challenging research \cite{janet2020accurate,jumper2021highly,avsec2021effective,graff2021accelerating,baek2021accurate,li2022competition,hsu2022learning,wang2023scientific}. Recent endeavors have ventured into utilizing AI techniques to enhance  quantum science. In the following, we introduce several seminal works in this context and highlight their relations and differences with our work.

\medskip
\noindent\textbf{Quantum state tomography.} AI-based quantum state tomography (AI-QST) \cite{torlai2018neural,cha2021attention,schmale2022efficient,koutny2022neural,wei2023neural} can be categorized into two primary groups based on the type of data processed by the neural networks. In the first category, researchers utilize a neural network, such as the restricted Boltzmann machine and deep Boltzmann machine, to represent the wave function of the target state, which takes as input a certain reference basis and outputs the corresponding density measurement \cite{torlai2018neural,schmale2022efficient}. In the second category, neural networks, such as multi-layer perceptrons and convolutional neural networks, are employed as a black box to infer the target density matrix from a set of measurement results \cite{xin2019local,lohani2020machine,palmieri2020experimental}.

In contrast to AI-QST methods, which primarily revolve around the design of neural networks for simulating quantum systems, our work places a distinct emphasis on the significance of data modalities in characterizing quantum systems. Furthermore, our approach focuses on learning the classical representation of a family of quantum states, as opposed to directly reconstructing the density matrix of a specific state, as done in previous studies. These distinctions enable a more efficient and practical solution for cross-platform verification.

\medskip

\noindent\textbf{Quantum system property prediction.} In addition to fully characterize quantum systems, AI is commonly employed to predict specific properties of these systems. These AI-driven methodologies harness the strong capacities of neural networks to derive partial information about quantum systems from a small amount of data when the quantum systems belong to a family of systems with similar structures, achieving high accuracy with fewer measurements and holding the potential to generalize to unknown quantum systems. Recent research has focused on predicting various properties of quantum systems using neural networks, including nonlocality detection \cite{deng2018machine,krivachy2020neural}, estimation of quantum state entanglement \cite{ma2018transforming, gray2018machine,rieger2023sample,koutny2023deep}, entropy estimation \cite{goldfeld2023quantum, shin2023estimating} and fidelity prediction \cite{zhang2021direct, wang2022quest, vadali2022quantum, wu2023quantum,du2023shadownet}. Despite these initial successes, using AI techniques for quantum cross-platform verification remains an open challenge.

\medskip
\noindent\textbf{Remark}. There are limitations when directly transferring these AI-based approaches to cross-platform verification.
Although theoretical evidence has shown that estimating fidelity between an ideal state and a mixed state can be achieved by polynomial measurements \cite{huang2020predicting}, it remains impractical to estimate the fidelity between two mixed states. 
When solely utilizing measurement data as the input of neural networks, their potential remains confined by the finite information encapsulated within the measurement outcomes. Alternatively, \cite{wang2022quest} takes the architectural features of quantum circuits as quantum system descriptors to predict the probability of successful trials (PST) \cite{tannu2019not}, underscoring the significance of circuit layout information in learning quantum states. In contrast to these methods which consider single modality, our work adopts a multi-modal learning paradigm that simultaneously incorporates measurement data and circuit data, leading to a reduced required number of measurements for cross-platform verification.

\section{The construction of a  class of quantum circuits towards near-term quantum machines}\label{app:prepare-state}
In this section, we will introduce the construction of a class of quantum circuits for near-term quantum machines. Considering the constraints of contemporary hardware and the inevitable quantum noise, we focus on a class of circuits that can be efficiently implemented on the devices. In the following, we start by discussing the design details of a class of quantum circuits tailored for cross-platform verification on near-term devices and elucidating the rationale behind our choice of these circuits. Afterward, we will introduce the different noise models employed in our experiments.

\subsection{Implementation of a class of circuits}\label{app:cir}

\begin{figure}[htp]
    \centering
    \includegraphics[width=0.4\linewidth]{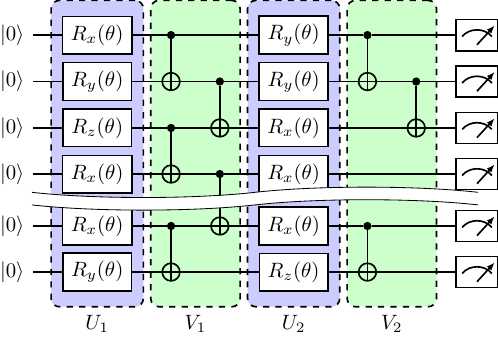}
    \caption{\small{\textbf{An example of HEA with $2$ layers in the ideal scenario.} Each layer consists of two blocks: a single-qubit gate rotation block and a two-qubit gate entanglement block. The single-qubit gates represent rotation gates that are efficiently supported by most quantum devices, while the two-qubit gates in the entanglement block connect only physically adjacent qubits.}}
    \label{fig:hea_layout}
\end{figure}

In the noisy intermediate-scale quantum (NISQ) era \cite{preskill2018quantum}, it is expected that the majority of near-term quantum algorithms will be tailored to align with the hardware constraints of near-term quantum devices. This alignment is critical to ensure that these algorithms remain practical and efficient for real-world application. Specifically, these circuits are tailor-made to leverage the available quantum operations offered by the hardware, such as the basis gates and the restricted set of effective two-qubit gates dictated by the qubit connectivity. In this way, this class of circuits mitigates the negative effects of device noise while making efficient use of the limited quantum resources. A diverse range of quantum algorithms are rooted in this circuit category, such as quantum neural networks \cite{havlivcek2019supervised,qian2022dilemma,du2023problem,du2022distributed,qian2022shuffle,tian2023recent}, variational quantum Eigen-solvers \cite{kandala2017hardware,tang2021qubit}, quantum simulations (Trotter methods or other high-order methods) \cite{smith2019simulating,barratt2021parallel,gibbs2022long,kim2023evidence}, and random circuit sampling \cite{arute2019quantum,zhu2022quantum}. Given their efficiency and versatile applications, we choose them as the quantum circuits for cross-platform verification in this paper. Following the general structure of circuits used in many near-term quantum algorithms, our implementation follows an alternating layered configuration, i.e.,

\begin{equation}\label{eq:app-hea}
    U = \prod_{l=1}^LU_l(\bm{\theta}_l),~\text{with}~ U_l=\prod_{n=1}^N e^{-i\theta_{l,n}H_{l,n}}V_l,
\end{equation}
where $\theta_{l,n}\sim \text{Unif}(0,2\pi)$ represents the angle of a rotation gate, $H_l\in\{\sigma_X, \sigma_Y, \sigma_Z\}$, $V_l\sim \mathbb{V}$ is an entanglement block consisting of multiple CNOT gates connecting physically neighboring qubits. An example of the circuit with $L=2$ is shown in Fig.~\ref{fig:hea_layout}.

The construction of a set of circuits following the form in Eq.~(\ref{eq:app-hea}) is as follows. Fixing the maximum layer $L$ and the system size $N$, we randomly generate a quantum circuit by randomly selecting $\theta_{l,n}$, ${H_l}$ and $V_l$. To be concrete, for each layer of the circuit, $\theta_{l,n}$ is determined by sampling from the distribution $\text{Unif}(0,2\pi)$ and ${H_l}$ is equally sampled from the set $\{\sigma_X, \sigma_Y, \sigma_Z\}$. For the entanglement block $V_l$, we randomly determine whether a CNOT gate exists between two neighboring qubits. Repeating this stochastic process allows us to generate multiple quantum circuits with diverse architectures, capturing the variability in circuit designs.

\subsection{Two quantum noise models}\label{app:cs}

\begin{figure}[htp]
    \centering
    \includegraphics[width=0.65\linewidth]{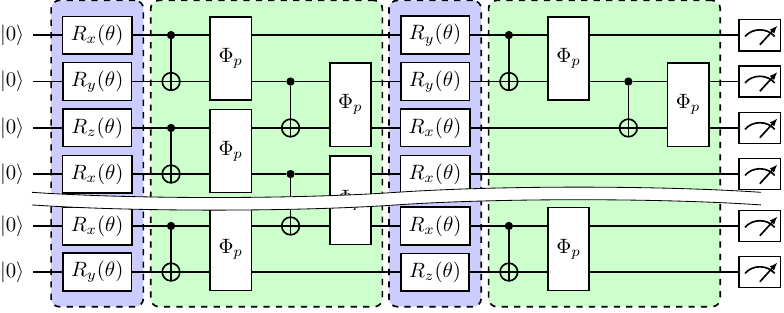}
    \caption{\small{\textbf{An example of the circuit with $2$ layers when exposed to two-qubit depolarizing noise.} The two-qubit depolarizing noise is modeled by applying a two-qubit depolarizing channel, denoted as $\Phi_p$, after each CNOT gate. The effect of this noise intensifies with an increase in the number of CNOT gates in the circuit.}}
    \label{fig:hea_layout_noisy}
\end{figure}

Due to the imperfections of quantum devices, the state $\rho$ prepared by the quantum circuit undergoes additional alterations. This process can be modeled by a device-relevant noise model $\mathcal{E}$ applied to the quantum state after the unitary transformation $U$: $\rho=\mathcal{E}(U^\dagger\rho_0U)$, where $\rho_0=(\ket{0}\bra{0})^{\otimes N}$ represents the initial state.
In our experiments, we introduce various noise models to comprehensively validate the performance of our model.
We first incorporate a two-qubit depolarizing channel after each two-qubit entanglement gate to simulate quantum noise as shown in Fig.~\ref{fig:hea_layout_noisy}, which is the primary source of noise in physical devices. The mathematical form of a two-qubit depolarizing channel is 
\begin{equation}
\Phi_p(\rho)=(1-p)\rho+p\frac{\mathbb{I}}{4},
\end{equation}
where $p$ is the strength of the depolarizing channel and ${\mathbb{I}}/{4}$ is the maximally mixed state for a two-qubit system. This approach enables us to subject circuits with varying numbers of entanglement gates to different levels of noise. Moreover, by adjusting the noise strength $p$ of the depolarizing channel, we can simulate various noise scenarios on different quantum hardware.

\begin{figure}[htp]
    \begin{minipage}[b]{0.45\textwidth}
        \centering
        \raisebox{-22mm}{
        \includegraphics[width=0.9\linewidth]{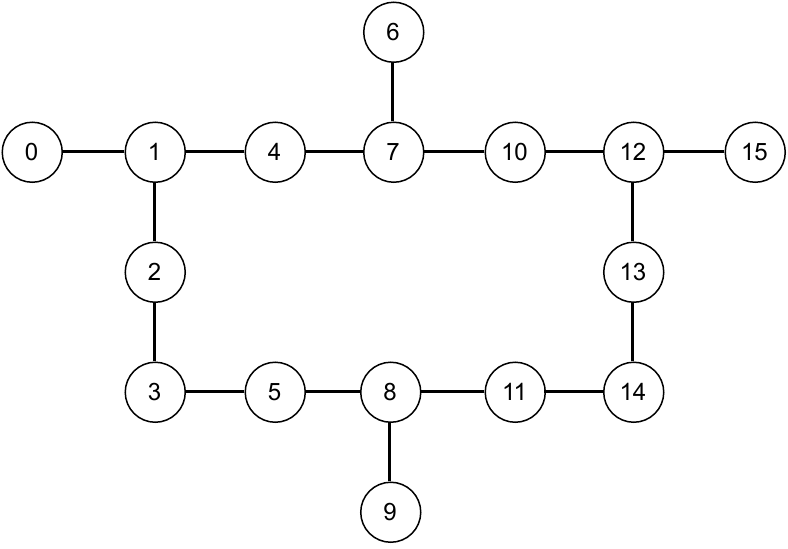}}
    \end{minipage}
    \hfill
    \begin{minipage}[b]{0.45\textwidth}
        \centering
        \raisebox{-22mm}{
        \includegraphics[width=0.7\linewidth]{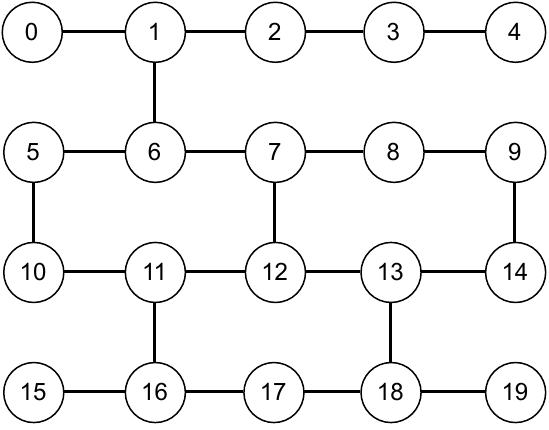}}
        \end{minipage}
    \caption{\small{\textbf{Comparison of the qubit connection map of two quantum devices \textit{ibmq\_guadalupe} (left panel) and \textit{ibmq\_boeblingen} (right panel) \cite{IBMquantum}.}}}
    \label{fig:device_topo}
\end{figure}

Another type of noise explored in our experiments is extracted from the real device noise. It encompasses a broader range of errors typical in near-term quantum devices, including factors like gate fidelity, qubit decoherence and relaxation times, and the specific chip topology of the device. In our numerical simulations, we acquire such noise models from four offline devices in IBM Quantum system. The qubit connection maps of a pair of quantum chips \textit{ibmq\_guadalupe} and \textit{ibmq\_boeblingen} are shown in Fig.~\ref{fig:device_topo}, which describes where a subset of qubits is physically connected to neighboring qubits, while others are not. These two devices display different patterns on the qubit connection.  Fig.~\ref{fig:device_para} records the device parameters that influence the fidelity of the prepared state. Unlike depolarizing noise, the noise associated with the \textit{ibmq\_guadalupe} device cannot be characterized by a single parameter. To be concrete, the values of T1 and T2 for a qubit determine the performance of CNOT gate, while a single error rate controls the remaining single-qubit gates. Additionally, we also consider the measurement error of each qubit, named as `\textit{prob\_meas0\_prep1}' and `\textit{prob\_meas1\_prep0}'. The value of \textit{prob\_meas0\_prep1} refers to the probability of returning $0$ when measuring on a prepared $\ket{1}$ state, while \textit{prob\_meas0\_prep1} refers to the probability of returning $1$ when measuring on a prepared $\ket{0}$ state. This level of detail provides a comprehensive noise model that more accurately reflects the complexities of real quantum hardware, which is essential for our cross-platform fidelity verification experiments.

\begin{figure}[htp]
    \begin{minipage}[b]{0.45\textwidth}
        \centering
        \begin{tabular}{cccc}
            \hline
            \textbf{Parameters} & \textbf{Min} & \textbf{Median} & \textbf{Max}\\
            \hline
            \textbf{T1}(us) & 38.89 & 71.86 & 119.14\\
            \textbf{T2}(us) & 14.55 & 88.30 & 142.49\\
            \textbf{RZ error} & 0 & 0 & 0\\
            \textbf{SX error} & 0.0002 & 0.0003 & 0.0018\\
            \textbf{$\sigma_X$ error} & 0.0002 & 0.0003 & 0.0018\\
            \textbf{CNOT error} & 0.0068 & 0.0101 & 0.0199\\
            \textbf{prob\_meas0\_prep1} & 0.0186 & 0.025 & 0.09\\
            \textbf{prob\_meas1\_prep0} & 0.0016 & 0.0057 & 0.0298\\
            \hline
        \end{tabular}
    \end{minipage}
    \hfill
    \begin{minipage}[b]{0.45\textwidth}
        \centering
        \begin{tabular}{cccc}
            \hline
            \textbf{Parameters} & \textbf{Min} & \textbf{Median} & \textbf{Max}\\
            \hline
            \textbf{T1}(us) & 24.11 & 73.76 & 98.44\\
            \textbf{T2}(us) & 9.60 & 79.06 & 153.18\\
            \textbf{u1 error} & 0 & 0 & 0\\
            \textbf{u2 error} & 0.0002 & 0.0005 & 0.0014\\
            \textbf{u3 error} & 0.0005 & 0.0010 & 0.0029\\
            \textbf{CNOT error} & 0.0061 & 0.0165 & 0.0277\\
            \textbf{prob\_meas0\_prep1} & 0.0202 & 0.0481 & 0.2128\\
            \textbf{prob\_meas1\_prep0} & 0.0062 & 0.0227 & 0.2128\\
            \hline
        \end{tabular}
    \end{minipage}
    \caption{\small{\textbf{Comparison of qubit and gate parameters of two quantum devices \textit{ibmq\_guadalupe} (left panel) and \textit{ibmq\_boeblingen} (right panel) \cite{IBMquantum}.} These two devices offer different sets of basis gates and various hardware errors. For \textit{ibmq\_guadalupe}, the supported basis gates include RZ, SX, PauliX, and CNOT, while for \textit{ibmq\_boeblingen}, they encompass u1, u2, u3, and CNOT.}}
    \label{fig:device_para}
\end{figure}

\section{Implementation details of MC-Net}\label{app:mc-net}

In this section, we elaborate on the implementation details of MC-Net. Recall the illustration in Fig.~\ref{fig:qsv}(c), the overall architecture of MC-Net comprises two branches, i.e., the measurement branch and the circuit branch, where each is responsible for handling distinct modality of data features. Namely, the measurement branch distills knowledge from sets of the measured bit strings; the circuit branch distills knowledge from the layout information of quantum circuits and the noise parameters associated with a specific quantum device. These two branches work in tandem to fuse the extracted features and learn a comprehensive representation of the quantum mixed state, ultimately facilitating the estimation of fidelity. In the subsequent sections, we will provide a detailed description of the dataset construction, and MC-Net's measurement branch and circuit branch architecture. Following that, we will introduce and compare several fusion strategies. Finally, we will present the training details.

\subsection{Dataset Construction}
We build the dataset as follows. The input $\bm{x}$ for MC-Net encompasses two distinct modalities: measurement results and circuit layout. For the measurement modality, we utilize the random measurement technique to obtain an efficient representation of a quantum state $\rho$. Specifically, we initially conduct random Pauli measurements on each qubit of $\rho$. Let $\ket{\bm{b}_{m}^{(s)}}=\ket{b_{m,1}^{(s)},...,b_{m,n}^{(s)},...,b_{m,N}^{(s)}}$ and $U_{m}^{(s)}=\otimes_{n=1}^NU_{m,n}^{(s)}$ be the binary outcomes and the corresponding measurement bases of the $m$-th measurements for the $s$-th sample. In the following explanation, we will omit the subscripts $m$ and $s$ for clarity. Next, we construct the classical shadow \cite{huang2020predicting} of the state $\rho$ based on the measurement data. For the $n$-th qubit, the shadow is expressed as $\hat{\rho}_n=3U_n^\dagger\ket{b_n}\bra{b_n}U_n-\mathbbm{I}$, which simultaneously encodes the information of both measurement basis and measurement outcomes. Following this, we flatten $\hat{\rho}_n$ into an eight-dimensional vector (each complex value is represented by two real values) and concatenate the shadows for all qubits, generating a feature vector $\bm{x}$ of length $8N$. Furthermore, to enhance the sensitivity of measurement outputs to the noise configuration, we can explicitly incorporate available noise-specific parameters into the feature vector $\bm{x}$. This process is repeated for $M$ times, resulting in a set of measurement vectors $\{\bm{x}_m\}_{m=1}^M$ that serve as the input for the measurement branch of MC-Net. For clarity, the subscript of $\bm{x}$ only specifies the $m$-th measured outcomes, omitting the device index.
 
To represent the device-specific circuit architecture, we encode the topology of a quantum circuit as a DAG, with device-specific noise information encoded into the node features. Following the convention in \cite{Qiskit, wang2022quest}, each node in the DAG corresponds to a quantum gate, and directed edges signify the temporal order of quantum gate execution. Additionally, two special node types, the input node, and the output node, represent the circuit's starting and ending points on a wire. For noisy devices simulated by depolarizing channels, we set the noise strength as an element of the feature vector of the CNOT gate node, while we set $0$ for others. For physical noise models, we consider the value of $T1$ and $T2$ of a qubit, as well as the gate and readout errors. (See SM~\ref{app:circuitnet} for a typical sample of encoding a circuit as a DAG).

\subsection{MC-Net}
MC-Net employs two branches, the measurement branch and the circuit branch, to separately distill the valid information that contributes to the verification of cross-platform from the two modalities.
To handle the measurement modality $\{\bm{x}_m\}_{m=1}^M$, we design a  convolutional neural network similar to  PointNet \cite{qi2017pointnet} as the measurement branch. Specifically, each measurement vector $\bm{x}_m$ is mapped to a higher-dimensional feature vector through shared convolutional kernels. Subsequently, we create a unified measurement feature vector by averaging over all $M$ feature vectors. This approach ensures that our model retains the permutation invariance of multiple measurements. 

For the circuit modality, we employ a graph neural network to extract the feature of the circuit-induced DAG. Following several layers of graph convolution, we perform average aggregation on all node features, resulting in a unified circuit feature vector. The measurement feature vector and circuit feature vector are then combined by low-rank bilinear pooling \cite{kim2016hadamard} to generate a unified representation of the quantum state (See SM~\ref{app:mc-net} for more details).

\subsubsection{Measurement Branch}
In MC-Net, the measurement branch independently processes the measurement data. As explained in the main text, recognizing that the measurement modality $\{\bm{x}_m\}_{m=1}^M$ collected from multiple snapshots is unordered, it is necessary that the measurement branch has the capability of preserving the permutation invariance of quantum measurements. To address this challenge, we draw inspiration from the successful application of PointNet \cite{qi2017pointnet} in handling point clouds in computer vision. In particular, we design a novel neural network termed `MeaNet' as the measurement branch of MC-Net, specifically designed to effectively capture the measurement features of a quantum state.

\begin{figure*}[htp]
\centering
\includegraphics[width=1.0\textwidth]{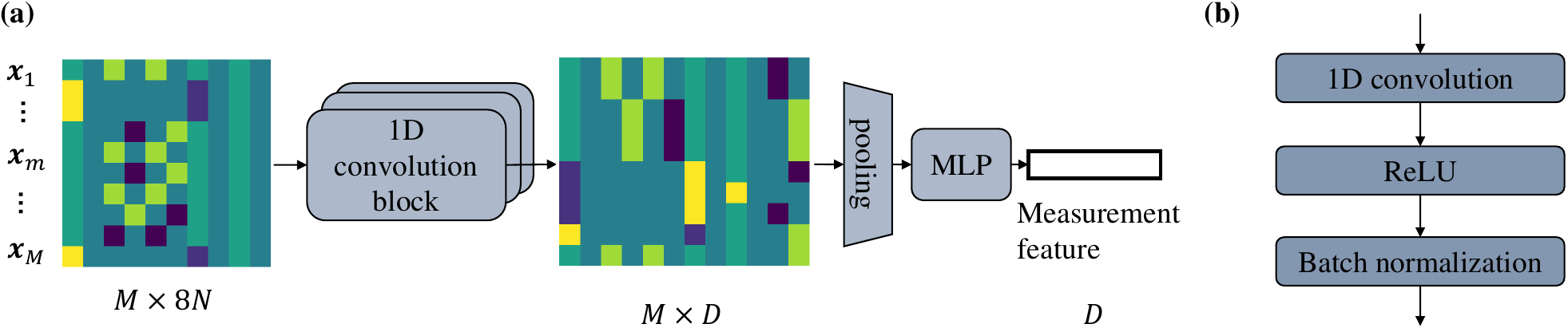}
\caption{\small{\textbf{Implementation of measurement branch of MC-Net.} (a) The architecture of MeaNet. The block `MLP' represents a multi-layer perceptron. The symbols `M', `N', and `D' refer to the number of measurements, the number of qubits, and the length of the output measurement vector, respectively. (b) The implementation of 1D convolution block in the MeaNet.}}
\label{fig:shadow-net}
\end{figure*}

The architecture of MeaNet is illustrated in Fig.~\ref{fig:shadow-net}. For each training example,  the network receives the modality of measurement in data features, i.e., a set of measurement vectors $\{\bm{x}_m\}_{m=1}^M$ generated from $M$ random Pauli measurements as introduced in SM. Initially, each original vector is projected to a higher-dimensional space using multiple 1D convolution blocks, and each of these blocks comprises several 1D convolutional kernels, a nonlinear activation function and a batch norm layer shown in Fig.~\ref{fig:shadow-net}(b). Let $\{\mathcal{K}_d\}_{d=1}^D$ denote the set of $D$ convolutional kernels. After applying convolution on a single measurement vector $\bm{x}_m$, we obtain a new vector $[\mathcal{K}_1(\bm{w}_1,\bm{x}_m), ..., \mathcal{K}_D(\bm{w}_D,\bm{x}_m)]$ with a length of $D$, where $\bm{w}_d$ is the weight of the $d$-th kernel. This process is repeated for all measurement vectors, yielding a set of measurement features $[[\mathcal{K}_1(\bm{w}_1,\bm{x}_1), ..., \mathcal{K}_D(\bm{w}_D,\bm{x}_1)],...,[\mathcal{K}_1(\bm{w}_1,\bm{x}_M), ..., \mathcal{K}_D(\bm{w}_D,\bm{x}_M)]]$ with the size $M\times D$. It is essential to highlight that each vector, corresponding to an individual measurement, shares the same convolution kernels. Subsequently, the features from convolution blocks are aggregated using average pooling followed by a multi-layer perceptron (MLP), resulting in a single vector with a length of $D$. This pooling operation ensures the preservation of measurement permutation invariance. The unified representation of a quantum state from multiple measurements is used for direct cross-platform fidelity estimation and can be integrated with additional information for further predictions.  

\subsubsection{Circuit Branch}\label{app:circuitnet}

The circuit branch of MC-Net is designed to process the circuit data. Unlike measurement data, quantum circuits can be better represented as structured graphs rather than vectors. As introduced above, we transform a quantum circuit into a Directed Acyclic Graph (DAG), where the edge and node connections precisely correspond to the circuit's layout and the execution order of quantum gates. To incorporate device-specific information into this DAG, we utilize a systematic rule for encoding device parameters as node features. We employ a Graph Neural Network (GNN), named `CircuitNet' to manage this circuit-induced DAG, enabling the model to learn the intrinsic impact of device noise on the final prepared state. In the following sections, we will first provide a rigorous explanation of the rule used to encode a quantum circuit as a DAG. Then, we will describe the design of the GNN in the circuit branch of MC-Net.

\begin{figure*}[htp]
\centering
\includegraphics[width=1.0\textwidth]{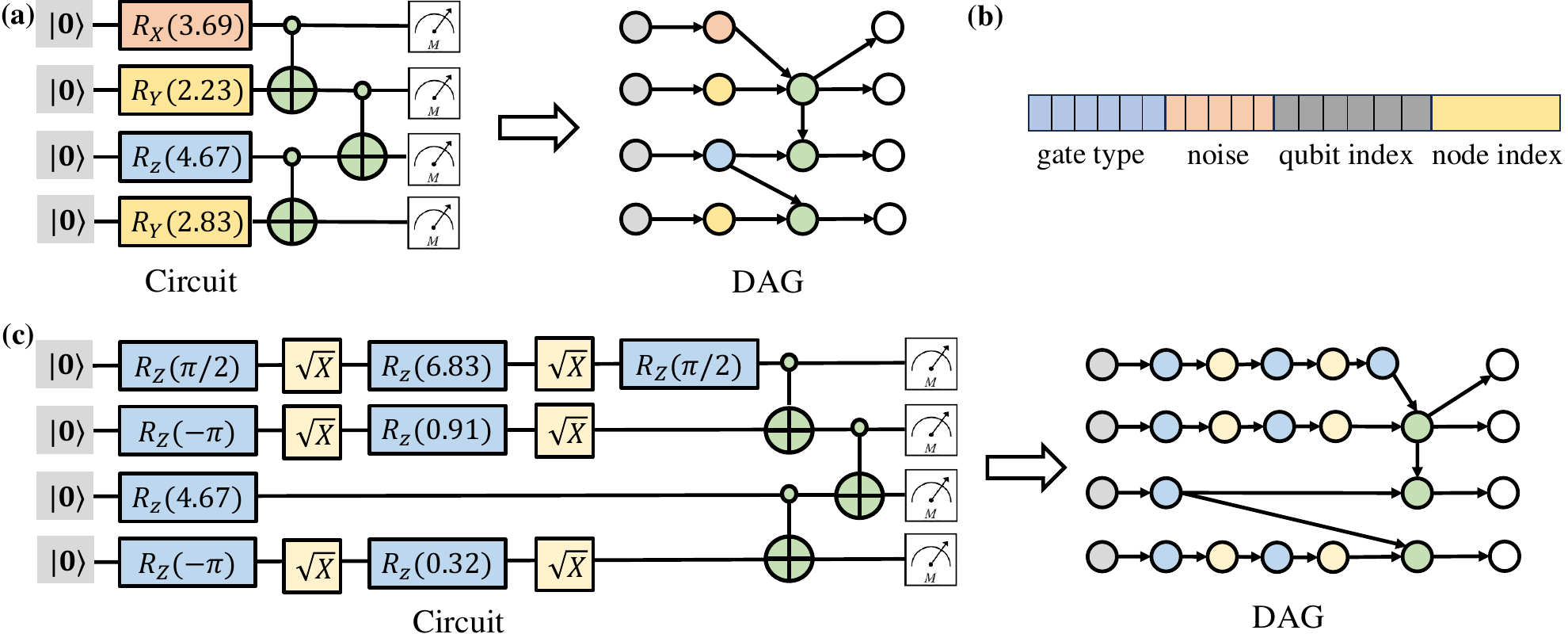}
\caption{\small{\textbf{Conversion from a quantum circuit to a DAG.} (a) An example of transforming an original quantum circuit into a DAG. (b) Construction of the node feature vector. (c) An example of transforming a quantum circuit transpiled for a platform into a DAG.}}
\label{fig:circuit-encode}
\end{figure*}

\noindent\textit{\underline{Directed Acyclic Graph reformulation.}} A typical example of encoding an original quantum circuit as a DAG is depicted in Fig.~\ref{fig:circuit-encode}(a). This representation encompasses six types of nodes: an input node, an output node, and four types of gate nodes. The input node (colored in grey) and the output node (colored in white) symbolize the initial state and measurement on a wire, respectively. The remaining nodes encompass three types of single-qubit rotation gates $\{R_X, R_Y, R_Z\}$ and a two-qubit CNOT gate. The directed edges that link neighboring quantum gates indicate the temporal order of quantum gate execution.

As shown in Fig.~\ref{fig:circuit-encode}(b), the feature vector of each graph node is composed of four key components: the type of the node, the noise features of the gate, the qubit index that the gate acts on and the node index. The type of the node is represented by a one-hot vector with a length equal to the number of gate types. In this representation, the $i$-th element is set to $1$, while the other elements remain $0$ to categorize the node into the appropriate type. The noise features of the gate depend on the specific device. For instance, in the case of numerical simulation with two-qubit depolarizing noise models, the noise feature of two-qubit gates is expressed as the strength parameter $p$ of the depolarizing channel, while the noise feature of other gates remains $0$. For real device noise models, the noise feature comprises additional information about each node, such as the values of $T1$ and $T2$ for the qubit, as well as gate and readout error rates. The qubit index is encoded as a one-hot vector, while the node index is encoded as an integer.

The construction rule of the DAG can be easily adjusted to accommodate the various basis gates supported by different quantum devices. When running a quantum circuit on a real quantum device that does not support the exact quantum gates used in the original circuit, it is necessary to transpile the circuit \cite{li2019tackling,nation2023suppressing,wilson2020just} to suppress the hardware noise. This ensures that the circuit is implemented using gates that are compatible with the capabilities of the target device.  As shown in Fig.~\ref{fig:circuit-encode}(c), the $R_X$ and $R_Y$ gates of the original circuit in Fig.~\ref{fig:circuit-encode}(a) are implemented with $R_Z$ and $\sqrt{X}$. While the original circuit and the transpiled circuit are theoretically equivalent in an ideal scenario, they can differ in practice due to variations in noise levels for different gate types in noisy conditions.  This discrepancy in noise characteristics and basis gates results in distinct DAGs for the two versions of the same circuit. This encoding scheme captures the device-specific variations of quantum circuits, allowing for meaningful distinctions between different quantum platforms.

\begin{figure*}[htp]
\centering
\includegraphics[width=0.75\textwidth]{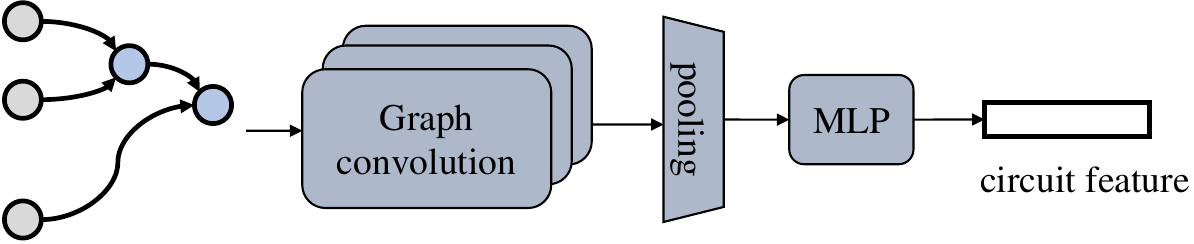}
\caption{\small{\textbf{The architecture of the circuit branch of MC-Net.} The block `MLP' represents a multi-layer perceptron.}}
\label{fig:circuit-net}
\end{figure*}

\medskip
\noindent\textit{\underline{Implementation of circuit branch.}} The architecture of the circuit branch of MC-Net is depicted in Fig.~\ref{fig:circuit-net}. Specifically, this branch takes the circuit-induced DAG as input and processes it using a graph convolution network. The graph convolution operation enables message passing among neighboring gates, effectively modeling the impact of gate errors in the quantum state evolution process. After several graph convolutions, we can obtain the circuit's feature vector by averaging the features of all nodes and projecting them into a higher-dimensional space using a multi-layer perceptron (MLP). This unified representation serves as a classical proxy for quantum circuits, ready for downstream tasks.

\subsubsection{Fusion module}\label{app:fuse}

Fusion plays a crucial role in the framework of multi-modal learning as it combines information from different modalities into a unified representation. In this context, consider two vectors: $\bm{x}\in \mathbb{R}^{D_1}$ and $\bm{y}\in \mathbb{R}^{D_2}$, representing feature of distinct modalities indexed by $1$ and $2$ respectively. The objective is to design a mapping $f: (\bm{x},\bm{y}) \rightarrow \bm{v}\in \mathbb{R}^{D}$ such that the fused feature $\bm{v}$ outperforms any individual unimodal data $\bm{x}$ or $\bm{y}$ in terms of performance of downstream tasks. Note that the dimensions of each feature vector ($D_1$, $D_2$, and $D$) need not be the same. Various fusion strategies for the selection of $f$ exist, including simple operation-based fusion, bilinear pooling-based fusion, and attention-based fusion, as discussed in \cite{zhang2020multimodal}.

\begin{figure*}[htp]
\centering
\includegraphics[width=0.9\textwidth]{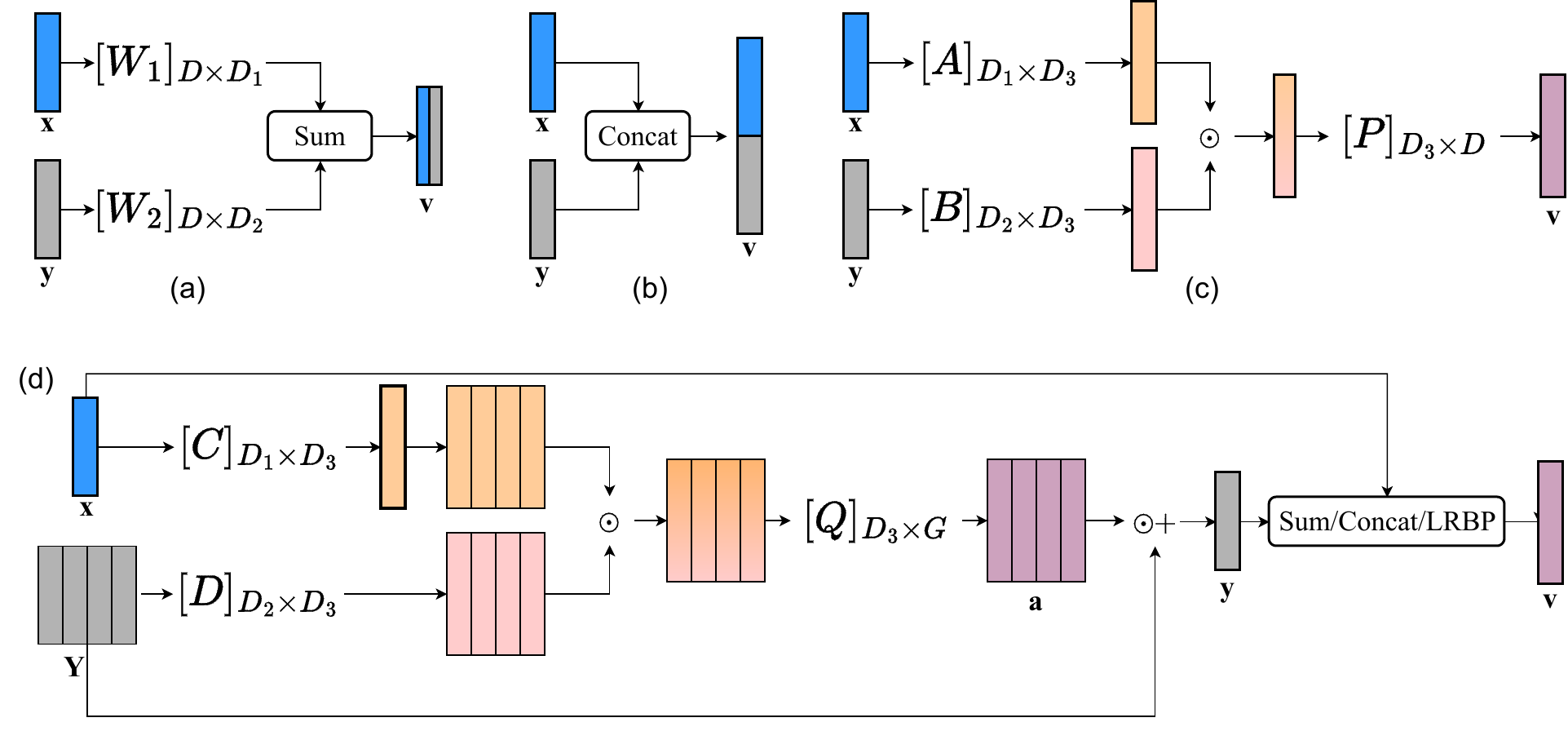}
\caption{\small{\textbf{Overview of various fusion strategies.} (a) Summation-based fusion. (b) Concatenation-based fusion. (c) Low-rank bilinear pooling-based fusion. (d) Attention-based fusion.}}
\label{fig:fuse-model}
\end{figure*}

\noindent\textbf{Simple operation-based fusion} \cite{nojavanasghari2016deep,vielzeuf2018centralnet,perez2019mfas}. In classical multi-modal learning, simple operations can be employed as the mapping function $f$, such as concatenation and summation. These operations have few or no trainable parameters and limited interactions of elements of different modality features.

\begin{itemize}
    \item Summation: As shown in Fig.~\ref{fig:fuse-model}(a), $\bm{v}=\bm{W}_1\bm{x}+\bm{W}_2\bm{y}\in\mathbb{R}^{D}$, where $\bm{W}_1\in \mathbb{R}^{D\times D_1}$ and $\bm{W}_2\in\mathbb{R}^{D\times D_2}$ are the weights that project $\bm{x}$ and $\bm{y}$ to the same dimension to satisfy the condition of element-wise addition.
    \item Concatenation: As shown in Fig.~\ref{fig:fuse-model}(b), $\bm{v}=[\bm{x}, \bm{y}] \in \mathbb{R}^{D_1+D_2}$.
\end{itemize}

\noindent\textbf{Bilinear pooling-based fusion.} Bilinear pooling involves calculating the outer product of two vectors, which can capture the complete set of interactions between every element of the feature vectors. Mathematically, the $i$-th element of the fused vector $\bm{v}$ after applying bilinear pooling on $\bm{x}$ and $\bm{y}$ can be expressed as
\begin{equation}
    \bm{v}_i = \bm{x}^T\bm{W}_i\bm{y},
\end{equation}
where $\bm{W}_i\in \mathbb{R}^{D_1\times D_2}$ is the $i$-th block of weight tensor $\bm{W}=[\bm{W}_1,...,\bm{W}_D]\in\mathbb{R}^{D_1\times D_2\times D}$. It is worth noting that for large values of $D_1$, $D_2$, and $D$, the computation cost of $\bm{W}$ can be quite high, which limits its practical feasibility.

Various works have been proposed to reduce the rank of the weight tensor $\bm{W}$ to restrict the number of parameters of $\bm{W}$. One typical approach \cite{pirsiavash2009bilinear}, called low-rank bilinear pooling (LRBP), is to decompose $\bm{W}_i$ into the product of two low-rank matrices: $\bm{W}_i=\bm{A}_i\bm{B}_i^T$, where $\bm{A}_i\in\mathbb{R}^{D_1\times D_3}$, $\bm{B}_i\in\mathbb{R}^{D_2\times D_3}$ and $D_3\leq \min(D_1,D_2)$. This allows us to rewrite $\bm{v}_i$ as:
\begin{equation}
    \bm{v}[i] = \bm{x}^T\bm{A}_i\bm{B}_i^T\bm{y}=\mathbbm{1}^T(\bm{A}_i^T\bm{x}\odot \bm{B}_i^T\bm{y}),
\end{equation}
where $\mathbbm{1}\in\mathbb{R}^{D_3}$ is a column vector of ones and $\odot$ represents Hadamard product. To further simplify the weight tensor $\bm{W}$, we use the same weight matrix $\bm{W}_i$ when calculating each element of the vector $\bm{v}$ and replace $\mathbbm{1}$ with a projection matrix $\bm{P}\in\mathbb{R}^{D_3\times D}$, as formulated by
\begin{equation}\label{eq:lrbp}
    \bm{v} = \bm{P}^T(\bm{A}^T\bm{x}\odot\bm{B}^T\bm{y}),
\end{equation}
where $\bm{A}\in\mathbb{R}^{D_1\times D_3}$, $\bm{B}\in\mathbb{R}^{D_2\times D_3}$. In this way, the number of parameters in the bilinear pooling-based fusion is reduced from $D_1\times D_2\times D$ to $D_3\times(D_1+D_2+D)$. The entire process of LRBP is shown in Fig.~\ref{fig:fuse-model}(c). In practice, a non-linear activation function is introduced to enhance the capacity of the fusion model \cite{kim2016hadamard}. To clarify, Eq.~(\ref{eq:lrbp}) is modified as follows:
\begin{equation}\label{eq:lrbp2}
    \bm{v} = \bm{P}^T(\sigma(\bm{A}^T\bm{x})\odot\sigma(\bm{B}^T\bm{y})),
\end{equation}
where $\sigma$ is a non-linear activation function, such as tanh in our work.

\medskip

\noindent\textbf{Attention-based fusion.} Attention mechanisms have been widely used in deep learning models, e.g. Transformer \cite{vaswani2017attention}, to help the models focus on the most relevant parts of the input data concerning the target. Typically, attention mechanisms involve estimating the attention distribution across elements in a set of feature vectors and then computing an attended vector based on this distribution. In the context of attention-based fusion, attention mechanisms are usually initially applied to unimodal features, and the attended features from different modalities are fused. Here, we combine attention mechanisms with low-rank bilinear pooling for feature fusion. Mathematically, the 
attention distribution is inferred as
\begin{equation}
    \bm{a} = {\rm softmax}(\bm{Q}^T(\sigma(\bm{C}^T\bm{x}\cdot \mathbbm{1}^T)\odot\sigma(\bm{D}^T\bm{Y}^T))),
\end{equation}
where $\bm{a}\in \mathbb{R}^{G\times S}$, $\bm{Q}\in \mathbb{R}^{D_3\times G}$, $\bm{C}\in \mathbb{R}^{D_1\times D_3}$, $\mathbbm{1}\in \mathbb{R}^S$, $\bm{D}\in \mathbb{R}^{D_2\times D_3}$, $\bm{Y}\in \mathbb{R}^{S\times D_2}$. For every glimpse $g\in[G]$, we calculate the attended feature vector $\bm{y}_g$ as a weighted sum:
\begin{equation}\label{eq:att-fea}
    \bm{y}_g = \sum_{s=1}^S\bm{a}_{s,g}\cdot \bm{Y}_{s,g}.
\end{equation}
The overall attended feature vector $\bm{y}$ is constructed by concatenating $\bm{y}_g$ in Eq.~(\ref{eq:att-fea}). The subsequent steps remain the same as those in the fusion of $\bm{x}$ and $\bm{y}$. The entire process is depicted in Fig.~\ref{fig:fuse-model}(d).

\subsection{Training details}

The training phase of MC-Net consists of two stages. The first stage involves separately training the measurement branch and the circuit branch of MC-Net. The second stage focuses on fine-tuning the entire MC-Net. With this two-stage training approach, MC-Net converges faster during training and achieves strong generalization on the test data.

\medskip
\noindent\textbf{Training of the measurement branch.} During the training of the measurement branch of MC-Net, the circuit branch and the fusion module are temporarily removed. This modified part is referred to as MeaNet. The feature vector outputted by MeaNet serves as the final classical representation $\bm{v}$ of a quantum state. Using the same objective function as described in Eq.~(\ref{eq:loss-func}) in the main text, we update the parameters of MeaNet with the Adam optimizer based on measurement data.

\medskip
\noindent\textbf{Training of the circuit branch.} Training of the circuit branch of MC-Net follows a similar procedure to that of the measurement branch. The measurement branch and the fusion module are temporarily removed, and the remaining part is named CircuitNet. The feature vector outputted by CircuitNet is considered as the final classical representation $\bm{v}$ of a quantum state. Using the same objective function as described in Eq.~(\ref{eq:loss-func}) in the main text, we update the parameters of CircuitNet with the Adam optimizer based on circuit data.

\medskip
\noindent\textbf{Fine-tune of MC-Net.} Once the MeaNet and CircuitNet are well trained, we begin to fine-tune the entire MC-Net, with its measurement branch and circuit branch inheriting parameters from MeaNet and CircuitNet, respectively. In comparison to the independent training of single branches, a smaller learning rate is used to update the parameters of the measurement branch, circuit branch, and fusion module. This fine-tuning process ensures that the different components of MC-Net work harmoniously and further improves the model's performance.

\section{Extension of MC-Net for cross-platform verification with multiple devices and the  prediction of other properties of quantum states}\label{app:extension}

In this section, we will explore the versatile extension of MC-Net for new quantum system learning tasks and illustrate this with two examples. To achieve this, we decouple the process of employing MC-Net for predicting cross-platform fidelity into two distinct components, allowing for later extensions. Subsequently, we will introduce methods for adapting MC-Net to perform cross-platform verification on multiple devices and for predicting quantum state purity.

The architecture of MC-Net can be divided into two main components: the network before the fusion module, denoted as $\mathcal{B}$, responsible for multi-modal learning to generate a general classical representation $\bm{v}$ of a quantum state; and the fusion module, denoted as $\mathcal{C}$, responsible for predicting the target of the task. With this framework, the predicted cross-platform fidelity $\mathcal{A}(\bm{x}_i, \bm{x}_j;\bm{w})$ in Eq.~(\ref{eq:risk}) can be derived as
\begin{equation}
    \mathcal{A}(\bm{x}_i, \bm{x}_j;\bm{w}=(\bm{w}_{\mathcal{B}},\bm{w}_{\mathcal{C}}))=\mathcal{C}(\bm{v}_i,\bm{v}_j;\bm{w}_{\mathcal{C}}), \text{with}\ \bm{v}_i=\mathcal{B}(\bm{x}_i;\bm{w}_{\mathcal{B}}),\bm{v}_j=\mathcal{B}(\bm{x}_j;\bm{w}_{\mathcal{B}}).
\end{equation}
Note that the network $\mathcal{B}$ is designed to be task-independent, whereas $\mathcal{C}$ is customized for specific tasks. By fixing $\mathcal{B}$ and adapting the task-specific module $\mathcal{C}$ to new tasks, we can apply MC-Net to various tasks.

\medskip
\noindent\textbf{K-device ($K\geq 2$) cross-platform verification.} K-device cross-platform verification involves simultaneous fidelity estimation among multiple states $\{\rho_i\}_{i=1}^K$. To achieve this, the network $\mathcal{B}$ is applied to multiple states to generate a set of representations $\{\bm{v}_i\}_{i=1}^K$. Subsequently, the average cross-platform fidelity over $K$ devices can be calculated as follows:
\begin{equation}
    \mathcal{F}(\{\bm{v}_i\}_{i=1}^K) = \mathcal{C}(\{\bm{v}_i\}_{i=1}^K;\bm{w}_{\mathcal{C}})=\frac{1}{K}\sum_{i=1}^K\frac{\left\langle\bm{v}_i, \bm{v}_j\right\rangle}{\left\|\bm{v}_i\right\|_2\left\|\bm{v}_j\right\|_2}.
\end{equation}

\medskip
\noindent\textbf{State purity estimation.} Quantum state purity is a measure of the degree of coherence and entanglement within a quantum state $\rho$, defined as
\begin{equation}
    \gamma = \Tr(\rho^2).
\end{equation}
To predict the purity of a quantum state, we employ a fully connected layer as the task-specific module $\mathcal{C}$. Let the weight matrix of the fully connected layer be denoted as $\bm{w}_\mathcal{C}$ with a shape of $(D\times 1)$, where $D$ is the length of the representation $\bm{v}$ of $\rho$ extracted by $\mathcal{B}$. Then the estimated state purity is
\begin{equation}
    \hat{\gamma}=\mathcal{C}(\bm{v};\bm{w}_{\mathcal{C}})=\bm{w}_{\mathcal{C}}^T\bm{v}.
\end{equation}
Other quantum system properties, such as state entanglement and entropy, can be inferred in the same manner in the framework of MC-Net.

\section{More numerical results}\label{app:more-results}
In this section, we present additional ablation studies on MC-Net. Specifically, we begin with the scaling behavior of MC-Net with respect to the system size. Following that,  we compare various aggregation strategies for the measurement data. Finally, we assess the performance of MC-Net using three different fusion models.

\subsection{Scaling behavior of MC-Net with respect to the system size}\label{app:scale-qubit}

Considering that the number of measurements required by classical shadow-based protocols scales exponentially with the number of qubits, we have experimentally investigated the scaling behavior of MC-Net in relation to the number of qubits to ascertain its resource efficiency compared to measurement-based protocols. We examine multiple system sizes ranging from 4 to 10 qubits. The test data for each system size are generated in the same manner as described in the main text. To ensure a fair comparison of performance across different system sizes, we employ the Relative Mean Square Error (RMSE) as our metric, defined as $\text{RMSE}=\frac{1}{S}\sum_s(\frac{\mathcal{F}^{(s)}-\hat{\mathcal{F}}^{(s)}}{\mathcal{F}^{(s)}})^2$.

\begin{figure*}[htp]
\centering
\includegraphics[width=0.45\textwidth]{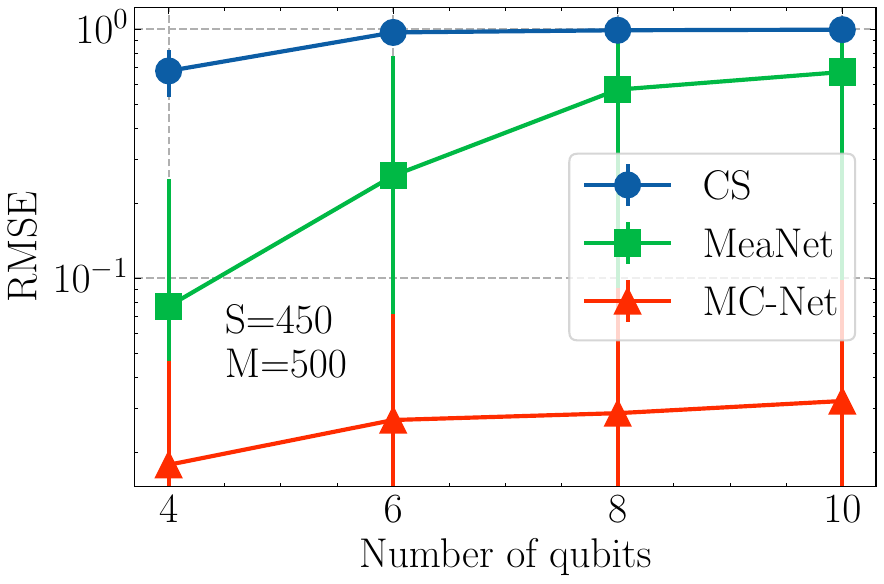}
\caption{\small{\textbf{The performance of classical shadow based cross-platform verification (CS), MeaNet and MC-Net concerning the number of qubits of the system.} The results are collected when fixing $S=450$ and $M=500$.}}
\label{fig:rmse_qubit}
\end{figure*}

The scaling behaviors of the Classical Shadow-based protocol (CS), MeaNet, and MC-Net methods, as the number of qubits increases from 4 to 10 when fixing $S=450$ and $M=500$, are depicted in Fig.~\ref{fig:rmse_qubit}. The CS method's RMSE begins at 0.680 and escalates dramatically to 0.996, signaling a decrease in precision with the rising qubit count. MeaNet, starting with a RMSE of 0.077, also sees a significant increase, culminating in a RMSE of 0.674 at 10 qubits. In stark contrast, MC-Net demonstrates remarkable stability and scalability. It starts with an impressively low RMSE of 0.018 and only marginally increases to 0.032 as the number of qubits reaches 10. This distinct scaling behavior highlights MC-Net's superior performance in managing larger quantum systems, showcasing a significantly more accurate performance than both CS and MeaNet and proving to be the most dependable method in preserving precision as the system's complexity increases. 

\subsection{Ablation study on the measurement data aggregation}

\begin{figure*}[htp]
\centering
\includegraphics[width=0.8\textwidth]{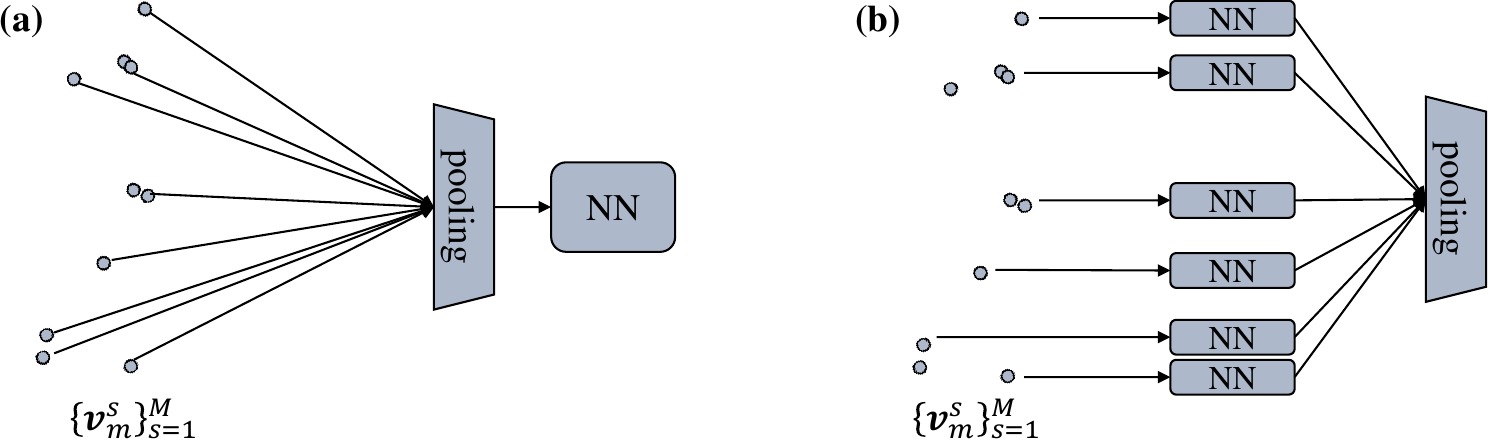}
\caption{\small{\textbf{Two strategies to process measurement modality.} The label `NN' is the abbreviation of neural networks. (a) Early average. (b) Lazy average.}}
\label{fig:avr-scheme}
\end{figure*}

We discuss the effect of the architecture of the MC-Net's measurement branch on the performance. Recall that the measurement modality after $M$ snapshots comprises a collection of $M$ measurement vectors $\{\bm{v}_m^t\}_{t=1}^M$. To process this format of measurement data, we introduce two strategies: early average and lazy average. In the early average strategy, we initially calculate the average measurement vector $\bm{\Bar{v}}=\frac{1}{M}\sum_{t=1}^M\bm{v}_m^t$. This averaged vector is subsequently fed into a neural network for further processing. This approach is advantageous in terms of memory and computational complexity, which remain at manageable levels even when dealing with a large number of measurements. However, it focuses solely on the statistics of multiple measurements, disregarding individual measurement results. Conversely, in the lazy average strategy, we regard each measurement vector, $\bm{v}_m^t$, as a point within the Hilbert space, offering a partial description of the quantum state. It is important to note that a sufficient number of these measurement vectors can collectively define a unique quantum state. With this respect, each measurement vector is individually processed by a neural network. Following this, we calculate the average over these processed measurement vectors to obtain the measurement feature. The schematic representation of these two schemes is depicted in Fig.~\ref{fig:avr-scheme}.

We examine the cross-platform fidelity prediction results of these two schemes on platforms with real device noise data, which is the same as the dataset introduced in the main text. To ensure a fair comparison, we maintain consistent numbers and shapes of weight matrices for each layer in both schemes. For the early average strategy, we selected a parameter shape for a 5-layer fully connected network as follows: $[(32, 64), (64, 128), (128, 256), (256, 256), (256, 256)]$. On the other hand, For lazy average strategy, the lazy average strategy involved $3$ layers of 1D convolution with the shape $[(32, 64), (64, 128), (128, 256)]$ prior to the pooling layer, and $2$ fully connected layers with shape $[(256, 256), (256, 256)]$ after pooling. Throughout the training process, both schemes shared the same hyper-parameters, including the learning rate, optimizer, and batch size, ensuring a consistent evaluation and comparison.

\begin{figure*}[htp]
\centering
\includegraphics[width=0.7\textwidth]{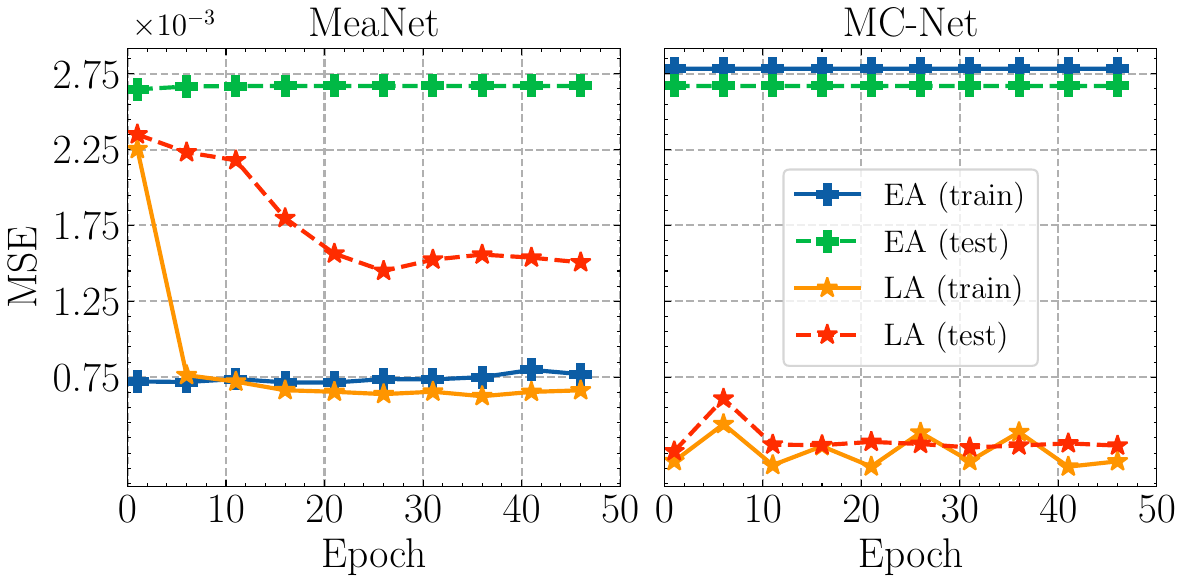}
\caption{\small{\textbf{The training and test performances of early average and lazy average for platforms with real device noise model.} The labels `EA' and `LA' are the abbreviations of early average and lazy average respectively. The left panel records the MSE reached by MeaNet during training. The right panel shows the MSE achieved by MC-Net.}}
\label{fig:ea_la}
\end{figure*}

The results are depicted in Fig.~\ref{fig:ea_la}. In the left panel, we observe the training and test performance of MeaNet using two strategies. While the early average and lazy average strategies yield similar MSE levels on the training dataset, the lazy average strategy results in a substantially lower MSE for the test data, specifically $1.5\times 10^{-3}$ compared to the $2.7\times 10^{-3}$ achieved by the early average. Analyzing the dynamic training process, it is evident that MeaNet with the early average strategy fails to exhibit effective training, with training loss and test performance remaining nearly constant. In contrast, the training loss and test error of MeaNet with the lazy average strategy show a notable early decrease followed by a smoother reduction. The right panel of Fig.~\ref{fig:ea_la} records the training and test behavior when applying two strategies to MC-Net. It demonstrates a similar trend for early average as observed in MeaNet. In the case of the lazy average strategy, it achieves a smaller training loss and test MSE than MeaNet, underlining the advantages of incorporating circuit layout data. These results emphasize that the lazy average strategy used in this study effectively captures essential measurement features for accurate cross-platform fidelity estimation.

\subsection{Comparison of fusion module}
The fusion module of MC-Net aims to integrate the measurement and circuit features into a unified state representation. This integration is expected to retain the essential information from both types of features, which significantly contribute to the specific task, while filtering out any unnecessary interference information present in each modality. To figure out the impact of various fusion methods on the final performance of MC-Net, we compare three kinds of fusion strategies: summation, concatenation, and low-rank bilinear pooling. In the summation-based and concatenation-based fusion methods, the trainable parameters are the weight matrix of fully-connected layers with shapes $(256,256)$ and $(512,512)$ respectively. For the low-rank bilinear pooling-based fusion, we set $D=D_1=D_2=D_3=256$ for the weight matrices $\bm{P}$, $\bm{A}$ and $\bm{B}$, as defined in Eq.~(\ref{eq:lrbp2}). The datasets used in this ablation study are the same as those described in the main text, including datasets with depolarizing noise and real device noise.

\begin{figure*}[htp]
\centering
\includegraphics[width=0.6\textwidth]{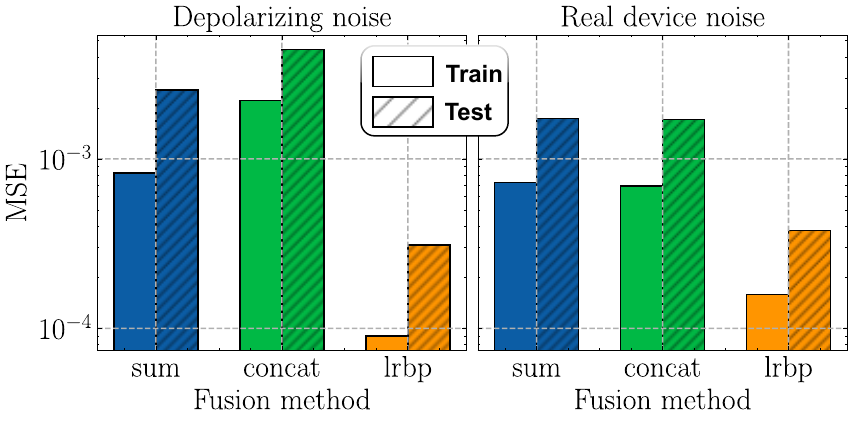}
\caption{\small{\textbf{Comparison of various fusion models in MC-Net.} The labels `sum', `concat', and `lrbp' represent the fusion methods of summation, concatenation, and low-rank bilinear pooling respectively. The left and right panels display the performance on datasets simulated with depolarizing noise and real device noise respectively.}}
\label{fig:fuse}
\end{figure*}

The numerical results are presented in Fig.~\ref{fig:fuse}. The left and right panels show the performance on datasets simulated with depolarizing noise and real device noise, respectively. It is evident that the low-rank bilinear pooling-based fusion significantly outperforms the other two fusion models. It achieves a one-order-of-magnitude reduction in both training and testing MSE on both datasets. While concatenating the two feature vectors might intuitively preserve more original information, the lack of sufficient interaction among the elements of the feature vectors leads to poorer performance in our task compared to low-rank bilinear pooling. These results validate the effectiveness of the fusion model in MC-Net and inspire us to design a more advanced fusion model to further enhance the performance of the multi-modal learning model.

\end{document}